\documentclass[12pt]{article}%
\usepackage{amssymb,amsmath,xcolor,graphicx,xspace,colortbl,ragged2e,rotating}
\usepackage{amsfonts}
\usepackage{amsmath}
\usepackage{amssymb}
\usepackage{amssymb,amsmath,xcolor,graphicx,xspace,colortbl,ragged2e,rotating}
\usepackage{boxedminipage}
\usepackage{caption}
\usepackage{graphics}
\usepackage{graphicx}
\usepackage{lineno}
\usepackage{mitpress}
\usepackage{ragged2e}
\usepackage{sectsty}
\usepackage{wrapfig}
\usepackage{xcolor}%
\providecommand{\U}[1]{\protect\rule{.1in}{.1in}}
\graphicspath{{ReconstructedQM-v4_graphics/}{ReconstructedQM-v4_tcache/}{ReconstructedQM-v4_gcache/}}
\DeclareGraphicsExtensions{.pdf,.eps,.ps,.png,.jpg,.jpeg}
\graphicspath{{ReconstructedQM-v2_graphics/}{ReconstructedQM-v2_tcache/}{ReconstructedQM-v2_gc
ache/}} \DeclareGraphicsExtensions{.pdf,.eps,.ps,.png,.jpg,.jpeg}
\providecommand{\U}[1]{\protect\rule{.1in}{.1in}} \textwidth=390pt
\subsectionfont{\normalfont\bf} \subsubsectionfont{\normalfont\itshape}
 \DeclareGraphicsExtensions{.pdf,.eps,.ps,.png,.jpg,.jpeg}
\providecommand{\U}[1]{\protect\rule{.1in}{.1in}}

\newdimen\dummy \dummy=\oddsidemargin 
\begin{document}

\title{Reconstructing Quantum Mechanics Without Foundational Problems }
\author{C. S. Unnikrishnan\\\relax\textit{Tata Institute of Fundamental Research}, \\\relax\textit{Homi Bhabha Road, Mumbai 400005, India} }
\maketitle

\begin{abstract}
I present a reconstruction of general Hamiltonian action mechanics that
eliminates all foundational problems of quantum mechanics. The key advance is
the completion of Hamiltonian mechanics to the universal mechanics of
particles based on action-waves, consistent with the inclusive validity of the
principle of stationary action. It is found that irreducible indeterminism is
intrinsic and universal at all scales of dynamics. The new action-wave
equation is the complete description of single dynamical histories, dissolving
the classical-quantum divide. The statistical theory of quantum mechanics
emerges as the ensemble average of modified action dynamics. The ensemble
average of the new action mechanics leads to a hybrid function consisting of
the action-waves and the probability density of the ensemble. This hybrid
wavefunction obeys the Schr{\"{o}}dinger equation, which is not a single
particle dynamical equation. The reconstructed mechanics without matter waves
is free of the cardinal problem known as the collapse of the wavefunction and
with that, the vexing issue of quantum measurement is resolved. Another
significant advance is the correct decoding of quantum entanglement and
purging of nonseparability and nonlocality in quantum correlations. The
action-waves do not carry the burden of divergent zero-point energy. The
reconstructed mechanics is in complete agreement with all empirical
requirements and in harmony with credible physical ontology.

\end{abstract}

\section*{Introduction}

Quantum mechanics is believed to be a universal theory of physical phenomena,
with a simple mathematical structure that is well established. Its empirical
success and reach are unprecedented. Yet, there is no consensus on a
satisfactory interpretation and understanding of quantum mechanics
~\cite{Feynman,Laloe,Weinberg}. The theory is beset with serious foundational
problems that have been discussed and debated for nearly a century.

In this paper, I reconsider the theoretical structure of \emph{quantum
mechanics} (QM) and find that it \emph{is factually an emergent theory,
obtained as the ensemble average of a more fundamental and universal theory of
single particle dynamics}. I present a reconstruction of quantum mechanics
(RQM) and the Schr{\"{o}}dinger equation, based on the new physical input of
the completion of the action mechanics of Hamilton \cite{Hamilton-2}. This is
not a re-interpretation of the Schr{\"{o}}dinger quantum mechanics; it is a
reconstruction requiring a fundamental conceptual change at the very
foundation of particle mechanics based on `carriers of action'. Thus, what we
conventionally call `classical mechanics' requires a fundamental modification
motivated by the universal validity of Hamilton's principle of stationary
action \cite{Hamilton-action}. The key finding is that instead of the
Hamiltonian action equation $\frac{\partial S}{\partial t}=-H$, the complete
and correct dynamical equation is $\frac{\partial\zeta(S)}{\partial t}%
=-\frac{i}{\varepsilon}H\zeta$, where the action-wave is $\zeta(x,t)=\exp
(iS(x,t)/\varepsilon)$ with $\varepsilon$ as a fundamental scale of action. It
will be shown that the action-wave is necessarily complex valued. The
action-wave mechanics already predicts that a fundamental irreducible
indeterminism is present universally in mechanics at all scales. The new
mechanics removes the distinction between classical and quantum mechanics, and
\emph{eliminates all foundational issues associated with quantum mechanics in
one sweep}. The vexing issues of the collapse of the wavefunction and the
quantum measurement problem are eliminated, while single particle interference
effects are correctly reproduced. Entanglement of particle states turns out to
be an ensemble notion. Correct quantum correlations without nonlocality and
nonseparability of particle states result from the local interference of action-waves.

Statistical average of the particle dynamics obeying the new action-wave
equation over the ensemble of dynamical histories results in a probability
density. This positive real probability density $\rho(x,t)=\chi\chi^{\ast}$
combined with the action-wave $\langle\zeta(x,t)\rangle$ gives a hybrid
ensemble function $\psi(x,t)=\left\vert \chi\right\vert \langle\zeta
(x,t)\rangle$, where $\chi=Ae^{i\phi}$. It is this ensemble function
$\psi(x,t)$ that obeys the Schr{\"{o}}dinger equation and exact Born's rule.
The Hilbert space structure with the square-integrability of this hybrid
complex function is emergent and refers to the ensemble of dynamical histories
and not to single histories. Thus, \emph{the Schr{\"{o}}dinger equation is not
an equation for single particle or single dynamical history. It is an ensemble
equation}. All previous attempts to understand the foundational structure of
quantum mechanics searched in vain for solutions in the analytical
interpretations of the Schr{\"{o}}dinger equation and its wavefunction.
Whereas, in fact, \emph{the complete resolution of the foundational problems
lay outside formal quantum mechanics, in the modified Hamiltonian action
dynamics, and not in the interpretation of the Schr\"{o}dinger equation or its
wavefunction}. It turns out that this is naturally accomplished without
changing the mathematical formalism and statistical structure of quantum
mechanics. Therefore, I am able to eliminate all foundational problems of
quantum mechanics without affecting its well tested statistical predictions.

The reconstruction of mechanics is based on identifying two distinct physical
entities in \emph{dynamics at all scales} -- the material particle, which is
the carrier of dynamical quantities like energy and momentum, and the
associated `action-waves' (or `phase-waves') that are the carriers of
`action', with no energy or momentum. The completion of Hamiltonian action
mechanics to a universal mechanics of particles based on action-waves does not
differentiate classical and quantum, relativistic and nonrelativistic,
microscopic and macroscopic etc. This is logically demanded by the physical
reason behind the universal applicability of the principle of stationary
action, namely that a wave-like entity capable of interference is present in
all dynamics. Therefore, the new Hamiltonian action-wave mechanics, rather
than quantum mechanics and the Schr{\"{o}}dinger equation, is the universal
basis of all mechanics. The particle's dynamics is linked to the phase of the
action-waves in a wave-particle connection. The existence of a fundamental
scale $\varepsilon$ of the action in the action-wave defines the fundamental
uncertainty in action, which reflects in all dynamics. This scale is
identified empirically as the Planck's constant $\hbar$.

I will show that \emph{both classical and quantum dynamics of a particle are
described by the same equation of evolution of the action-wave},
$\frac{\partial\zeta(S)}{\partial t}=-\frac{i}{\varepsilon}H\zeta$. Hamilton's
action equation $\frac{\partial S}{\partial t}=-H$ (called the Hamilton-Jacobi
equation \cite{Hamilton-history}) descends from this master wave-equation for
dynamics. \emph{Material particles do not behave as waves or superpose; there
are no matter waves}. Quantum mechanics is factually the ensemble average of
the new Hamiltonian action mechanics. The \emph{wavefunction of the
Schr{\"{o}}dinger dynamics is really a hybrid entity} involving the ensemble
averaged probability density for the material particle, without superposition,
and action-waves of quantum possibilities that can superpose and interfere
event by event. The Schr\"{o}dinger wavefunction does not pertain to single
quantum history, as believed hitherto, but to the ensemble average of such
histories. Thus, both the probabilistic particle dynamics and the vital
quantum interference are intact. There is a radical change in the physical
paradigm and interpretation, while the mathematical super-structure and
operations remain intact because the quantum mechanical calculations pertain
to the averages over the (virtual) ensemble. This explains all features of
quantum mechanics.\emph{\ }

\section{Foundational Problems of QM}

The standard theory of quantum mechanics is based on the Schr{\"{o}}dinger
equation for the time evolution of a wavefunction, which refers to single
dynamical history from preparation to observation, in all interpretations. The
wavefunction is a single functional dependent on the coordinates of the
particle, in the case of a single particle, and on the 3N coordinates for N
particles. Though the concept evolved from de Broglie's matter-waves, the
multiparticle wavefunction evades consistent physical interpretations.
However, all standard approaches treat the wavefunction as having a space-time
support because only then one can understand how local interactions in space
and time can alter the wavefunction. Also, the entire notion that quantum
mechanics has \textit{nonlocal} features is dependent on changes that happen
to the wavefunction across spatially separated regions.

The characteristic and essential trait of quantum mechanics is the inherent
indeterminism; the theory has only probabilistic predictions, despite being
described by the deterministic Schr{\"{o}}dinger equation of evolution. This
indeterminism is linked to the wave-particle duality and the uncertainty
principle, encoded mathematically in the commutation relations between
observables. Indeterminism results in different and random outcomes in
measurements or observations on the quantum states described by the same
wavefunction. In practise, this is manifest in the statistically distributed
results on identically prepared ensemble of quantum systems, represented by
the same wavefunction or the quantum state. As an example, in the physical
state of an electron represented by $|z+\rangle$, its spin component is known
with \emph{certainty} to be in the positive z direction. When measured along
the direction $z^{\prime}\text{,}$ such that $z\cdot z^{\prime}=\cos\theta$,
the \emph{probability} for observing the electron in the state $|z^{\prime
}+\rangle$ is $cos^{2}(\theta/2)$. Each measurement on the identical states
$|z+\rangle$ then returns \textit{random results} of $|z^{\prime}+\rangle$ and
$|z^{\prime}-\rangle$, with a fraction $cos^{2}(\theta/2)$ in the $+$ direction.

The conventional formalism of QM leads to several vexing problems, debated as
its `birth defects'~\cite{Laloe,Dipanker,Norsen}. Foremost is the core issue
known as the \textit{collapse of the wavefunction during an observation}. The
quantum physical state is represented by a state vector in Hilbert space,
$|\psi\rangle$. This can be linear superposition of component states
$|\psi\rangle=\sum c_{i}|\psi_{i}\rangle$, with $\sum\left\vert c_{i}%
\right\vert ^{2}=1$. The coefficients $c_{j}$ determine the probability
$p_{j}$ to be observed in the particular state $|\psi_{j}\rangle$ through
Born's rule $p_{j}=\left\vert c_{j}\right\vert ^{2}$, with $\sum\left\vert
c_{j}\right\vert ^{2}=1$. An unknown general state could be represented
similarly with values of $c_{i}$ unspecified. When a measurement is made, only
one (or a subset) of the possible results $i=n$ materializes, stochastically.
The quantum state is said to \emph{collapse} or \emph{reduce} uniquely to
$|\psi_{n}\rangle$, with \emph{other components disappearing instantaneously}.
Since Born's rule dictates that $\left\vert c_{n}\right\vert ^{2}=1$ after
observing the system in the state $|\psi_{n}\rangle$, consistency demands the
instantaneous vanishing of all other components. Therefore, any space-time
representation of the wavefunction implies a nonlocal reduction of some entity
in space and time. On the other hand, the interpretation of the wavefunction
as a purely mathematical entity to represent the state of the particle in the
abstract Hilbert space, without a space-time counterpart, fails in the
physical explanation of even simple interference experiments involving local
interactions (as illustrated in the next section).

The problem of collapse gets more difficult for multi-particle systems. For a
two-particle two-state system, with possible states $|+\rangle$ and
$|-\rangle$, the joint state may be a linear superposition $\psi_{e}%
=c_{a}|+_{1}\rangle|-_{2}\rangle+c_{b}|-_{1}\rangle|+_{2}\rangle$, called an
entangled state. This has no quantum mechanical description as the joint state
(product state) of two independent quantum systems, $|s_{1}\rangle
|s_{2}\rangle$. The most general states of the particles (1 and 2) with two
base-states are $|s_{1}\rangle=a_{1}|+_{1}\rangle+b_{1}|-_{1}\rangle$ and
$|s_{2}\rangle=a_{2}|+_{2}\rangle+b_{2}|-_{2}\rangle\text{.}$ Then the most
general product state is
\begin{equation}
\psi_{1,2}=(a_{1}|+_{1}\rangle+b_{1}|-_{1}\rangle)\otimes\left(  a_{2}%
|+_{2}\rangle+b_{2}|-_{2}\rangle\right)
\end{equation}
But the state $\psi_{e}$ cannot be written as $\psi_{1,2}$. Then,
\textit{there is no definite quantum mechanical state for either particle}!
Each particle is not in the state $|+\rangle$, not in the state $|+\rangle$,
or not in a general superposition $a|+\rangle+b|-\rangle$. The particles have
their quantum mechanical existence only as a single non-separable system.
However, each particle is available to the experimenter as separate systems,
possibly in widely separated laboratories. Measurement on one particle returns
a definite value $+$ or $-$ and a corresponding quantum mechanical state, say
$|-_{1}\rangle$. This means that $\psi_{e}$ has collapsed to the state
$|-_{1}\rangle|+_{2}\rangle$. Then, instantaneously and nonlocally, the other
particle acquires a definite new state $|+_{2}\rangle$ without any interaction
or observation. \emph{An observation on one particle has created individual
states for both, from a situation in which neither had quantum state}. The
conceptual difficulty arising due to this induced nonlocal collapse is the
basis of the Einstein-Podolsky-Rosen (EPR) discussion \cite{EPR} and the
results about the conflict between quantum mechanics and Einstein locality.
(EPR assumed Einstein locality and that the physical state of matter could not
be changed nonlocally from a spatially separated region. Then they concluded
that quantum mechanics was not complete -- wavefunction description is not in
one-to-one correspondence with the physical state, because the wavefunction of
a particle can be changed by a measurement on another particle. See appendix 4
for details).

The ensuing \emph{quantum measurement problem}~\cite{Norsen,Leggett} is
closely linked to the collapse of the wavefunction. If two physical states of
a system are represented as $|1\rangle_{s}$ and $|2\rangle_{s}$ the physical
system can also be in the state $a|1\rangle_{s}+b|2\rangle_{s}$ \ with
$|a|^{2}+|b|^{2}=1.$ During a measurement, another physical system acting as
the ideal measurement `apparatus' with pointer states $|1\rangle_{A}$ and
$|2\rangle_{A}$ interacts with the system and forms the correlated and
entangled state
\begin{equation}
\left\vert S_{sA}\right\rangle =a|1\rangle_{s}|1\rangle_{A}+b|2\rangle
_{s}|2\rangle_{A}%
\end{equation}
Then, \emph{neither the microscopic system nor the macroscopic apparatus has
any individual physical state}. The bizarre situation is unavoidable in
standard QM irrespective of the size and mass of the `apparatus'. It should be
understood correctly that neither of the physical systems is in any
superposition of the two allowed states; each system has no physical state of
its own within QM \cite{Schrod-Ent} (this crucial point is often not
adequately understood, diluting \ the severity of the problem). This is the
same as the much discussed problem of the Schr\"{o}dinger's cat, where the
`apparatus system' is macroscopic and possibly living \cite{Unni-SCat}%
.\footnote{In this general review paper written in 2005, I had already
indicated the empirical factors that hold the key to eliminating the
foundational issues in QM. That the theoretical advance needs modifying
Hamilton's action mechanics was not expected then.} The final (conscious)
experience about the measurement is, however, either the state $|1\rangle
_{s}|1\rangle_{A}$ or the state $|2\rangle_{s}|2\rangle_{A}$. This is the
collapse of a nonseparable entangled state into a joint product state; it is
the collapse of the superposition of multi-particle states to one specific
separable product state in which each system now has a definite individual
state. The process by which this `state reduction' and the appearance of a
unique pointer state happen is a mystery.

The quantum measurement problem is more severe than it looks in the case of
\ the `system and apparatus'. We saw that there is no QM state for the
apparatus \emph{until observed}. However, to observe the apparatus, an
observer $O\text{{}}$ is required, which might be a more complex apparatus or
a human being, eventually. But, if QM is universally applicable, all that can
happen is a more complex entanglement
\begin{equation}
\left\vert S_{sAO}\right\rangle =a|1\rangle_{s}|1\rangle_{A}\left\vert
1\right\rangle _{O}+b|2\rangle_{s}|2\rangle_{A}\left\vert 2\right\rangle _{O}%
\end{equation}
and not a collapse to any definite state for the observer, or the apparatus,
or the system. It is an infinite regress, until an agent that does not obey
the law of linear superposition in quantum mechanics is artificially
introduced. This is what is done by the standard interpretation, when it
assumes a `classical apparatus' for measurements. Desperate desire for the
resolution of the quantum measurement problem has prompted some physicists to
take the extreme speculative view that an undefined entity called
`consciousness' is involved in breaking the system-apparatus entanglement.
Such proposals are inconsistent because they invoke structures built out of
matter that do not obey the laws of quantum mechanics.

Finally, despite many proposals, there is no satisfactory understanding of the
emergence of the classical macroscopic world from what is believed to be the
more fundamental quantum world. Are there two separate physical worlds with a
transition zone, with a fundamental quantity like mass as the parameter, or is
dynamics based on a universal principle and single evolution equation? The
answer is in the solution to the measurement problem.

\emph{The Reconstructed Quantum Mechanics (RQM), derived from Reconstructed
Hamiltonian Action Mechanics (RHAM), is free from all these foundational
problems}.

\section{Reconstructing Hamiltonian Mechanics}

\begin{figure}[ptb]%
\centering
\includegraphics[
natheight=2.221200in,
natwidth=4.597100in,
height=1.6139in,
width=3.3176in
]%
{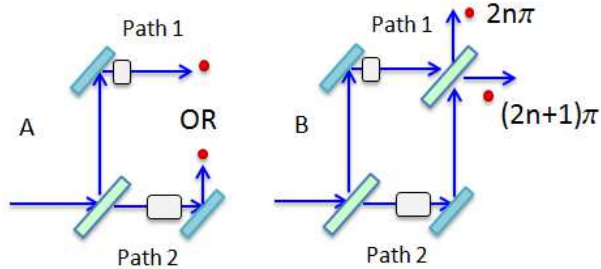}%
\caption{Two essential empirical features: A) a particle is always detected as
a whole and any interpretation that requires a `split' in space before
detection violates fundamental conservation laws. B) In a situation of
interference, the outcome can be deterministically controlled, particle by
particle, by the combined local physical processes in each path. This requires
the presence of a physical entity in both paths. }%
\label{Empirical}%
\end{figure}

There are two essential empirical facts in microscopic dynamics, represented
in figure \ref{Empirical}, that are important for the correct physical theory.
Quantum particle dynamics is often about multiple possibilities that can
interfere. For such a situation, the spatial paths could be well separated.
What is clear is that the particle cannot be in both paths without violating
all fundamental conservation laws because all experimental results are as if
the particle remains as a single whole at all times, with its intrinsic
properties (mass, electric charge, spin etc.) not divided or smeared out.
Therefore, a virtual copy would violate the conservation laws and division of
any sort would not be consistent with the general empirical results. However,
the interference phenomena, characteristic of quantum dynamics, demand that
there is some entity that \emph{divides and be physically present in the
multiple paths}.%
%TCIMACRO{\FRAME{ftbpFU}{3.3176in}{1.6139in}{0pt}{\Qcb{Two essential empirical
%features: A) a particle is always detected as a whole and any interpretation
%that requires a `split' in space before detection violates fundamental
%conservation laws. B) In a situation of interference, the outcome can be
%deterministically controlled, particle by particle, by the combined local
%physical processes in each path. This requires the presence of a physical
%entity in both paths. }}{\Qlb{Empirical}}{empirical.png}%
%width 3.3176in;  height 1.6139in;  depth 0pt;  original-width 4.5971in;
%original-height 2.2212in;  cropleft "0";  croptop "1";  cropright "1";
%cropbottom "0";  filename 'Empirical.png';file-properties "XNPEU";}} }%
%BeginExpansion
%EndExpansion

In a Mach-Zehnder interferometer, the relative phase in the two paths
determines the probability for the particle to exit through either of the
output ports. When the phase difference is $2n\pi\text{,}$ \emph{every
particle} will exit via one port and for $\left(  2n+1\right)  \pi$, via the
other port, \emph{deterministically, particle by particle}. However, the total
phase difference depends on independent phase shifts in the two paths, due to
the positions of the mirrors, local phase shifters etc. Therefore, it is
obvious and unavoidable that for each particle there is \emph{some non-matter
physical entity in both paths}, which can account for the different and
independent phase shifts that finally combine to give the phase difference
that determines the port of exit.

The important new finding is that taking care of these two empirically
essential features provides a unique reconstruction of quantum mechanics that
is free of all the foundational problems. However, this reconstruction has to
start from a reconstruction of classical Hamiltonian mechanics itself, because
it turns out that \emph{classical mechanics in the present form is not
complete}. When W. R. Hamilton introduced the principle of stationary action
in 1832, it was generalised from Fermat's least action principle for optics
\cite{Hamilton-action}. He developed a universal and \textquotedblleft general
method for expressing the paths of light, and of the planets\textquotedblright%
. Hamilton also discussed the rationale involved in the principle, through
Huygens' theory of wave optics. In the subsequent papers
\cite{Hamilton-2,Hamilton-1} that described the method for mechanics based on
the `characteristic function', or the action function $S(x,t)$, he wrote the
general equation for dynamics as
\begin{equation}
\frac{\partial S}{\partial t}=-H,\text{\quad}H=T+V
\end{equation}
The momentum is derived from the action as $\partial S/\partial x=p$. The
equation was called the Hamiltonian equation by C. G. J. Jacobi, who developed
it further mathematically \cite{Hamilton-history}. However, this equation does
not explicitly acknowledge the \emph{action as a central property residing in
a wave-like entity} and does not represent the core reason for the universal
validity of the principle of stationary action! The principle is operative
because action is physically manifest in a wave property where the phase of
the wave is the scaled action. Hence, I complete Hamilton's action mechanics
by modifying the equation to a new universal form
\begin{equation}
\frac{\partial\zeta(S)}{\partial t}=-\frac{i}{\varepsilon}H\zeta
\label{Ham-actionwave}%
\end{equation}
where the function $\zeta(x,t)=\exp(iS(x,t)/\varepsilon)$. I call this
function the `Action-wave', since it has a direct physical presence. This
equation is the key step in the correct description of single particle,
single\ dynamical history in the reconstructed universal dynamics. The
parameter $\varepsilon$ is the fundamental scale of action that is to be
determined empirically. This equation contains Hamilton's equation $\partial
S/\partial t=-H$.

The first feature I note is that the action-wave is necessarily complex
valued. For, the relation between the first order time evolution and the
Hamiltonian involves second order spatial derivatives. Therefore, the
dynamical equation cannot be built on a real periodic function like
$\cos(S/\varepsilon)$ or on a real linear combination of sinusoidal functions;
the action wave needs both quadratures as a complex valued function,
$\cos(S/\varepsilon)+i\sin(S/\varepsilon)$. \ Therefore, the fundamental
equation for dynamics involves complex numbers because complex valued action
waves are the basis of dynamics at all scales. This feature is then
\emph{inherited} by quantum mechanics. The action-wave has unit amplitude,
dictated by the dynamical equations $\partial S/\partial t=-H$ and $\partial
S/\partial x=p$.

\begin{figure}[ptb]%
\centering
\includegraphics[
natheight=2.056700in,
natwidth=6.877000in,
height=1.3507in,
width=4.4682in
]%
{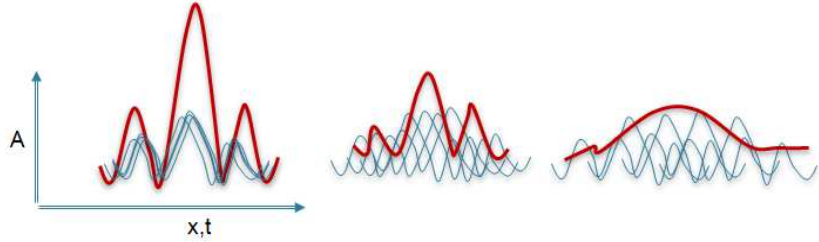}%
\caption{Schematic diagram of how the dephasing of action-waves linearly
manifests as a dissipationless diffusion of the phase and the total amplitude
of the action-waves.}%
\label{dephasing}%
\end{figure}

We note the immediate main result that the new Hamiltonian action-wave
equation predicts \emph{intrinsic indeterminism in dynamics at all scales}.
With $H=\left(  p^{2}/2m\right)  +V$, and $p=\partial S/\partial x$ we get,
\begin{equation}
i\varepsilon\frac{\partial\zeta(S)}{\partial t}=\left(  -\zeta\frac{\partial
S}{\partial t}\right)  =\left(  -\frac{\varepsilon^{2}}{2m}\nabla
^{2}+V\right)  \zeta=\zeta\left(  \frac{p^{2}}{2m}+V\right)  -\zeta
\frac{i\varepsilon}{2m}\nabla^{2}S
\end{equation}
In addition to the familiar Hamiltonian mechanics in the first term, there is
an additional term proportional to $\varepsilon$ and to the
\emph{concentration of the action in space}, $\nabla^{2}S$. The equation is
the single particle, single dynamical history equation. Hence, this term
points to the variations in an ensemble of action-waves, for the dynamics of a
single particle. \emph{When }$\varepsilon$\emph{\ is tiny, this term is
negligible. But it will become significant for microscopic dynamics}. To see
this clearly, consider the situation when $p\simeq0$. Then, we get
\begin{equation}
\frac{\partial S}{\partial t}=\frac{i\varepsilon}{2m}\nabla^{2}S
\end{equation}
This resembles the diffusion equation, but the diffusion constant is pure
imaginary. Therefore, this is not dissipative diffusion, but
\textit{dephasing} of the action waves (fig. \ref{dephasing}). The action can
be specified only to the precision of $\varepsilon$; thus the fundamental
uncertainty in action is the quantity $\varepsilon\text{.}$ Therefore,
dynamics at all scales have the \emph{same} unavoidable intrinsic uncertainty
in action. This is empirically determined as the Planck's constant,
$\varepsilon\equiv\hbar\simeq1.055\times10^{-34}$ J-s, which is a miniscule
amount of action. We have found that the fundamental uncertainty is not
limited to microscopic world and quantum dynamics. It is universal, but
negligible for macroscopic dynamics. This resolves the primary fundamental
puzzle discussed in the context of quantum mechanics, namely the origin of
indeterminism. Its source is action-wave dynamics, and it is universal.
Classical mechanics is not deterministic. Classical mechanics is dynamics in
which the irreducible indeterminism is negligible. Hence, we have also
answered another foundational question, about a quantum-classical divide and a
transition domain. There is no such divide. \emph{The new dynamics predicts
that the experiments that search for evidence of a transition, either at
Planck mass scale or any other scale in mesoscopic systems, will have a
definite null result}.%
%TCIMACRO{\FRAME{ftbpFU}{4.4682in}{1.3507in}{0pt}{\Qcb{Schematic diagram of how
%the dephasing of action-waves linearly manifests as a dissipationless
%diffusion of the phase and the total amplitude of the action-waves.}%
%}{\Qlb{dephasing}}{dephase.png}{\special{ language "Scientific Word";
%type "GRAPHIC";  maintain-aspect-ratio TRUE;  display "USEDEF";
%valid_file "F";  width 4.4682in;  height 1.3507in;  depth 0pt;
%original-width 6.877in;  original-height 2.0567in;  cropleft "0";
%filename 'Dephase.png';file-properties "XNPEU";}} }%
%BeginExpansion
%EndExpansion

Figure \ref{dephasing} shows how the dephasing of the action waves translates
to the diffusion of the amplitude of the waves. The effect is non-dissipative
and progresses to reduce the concentration of action, $\nabla^{2}S$. This
manifests in the linear diffusion of the amplitude of the action-waves, which
is of crucial importance when we discuss the reconstruction of quantum
mechanics later.

Surprisingly, \emph{with the Reconstructed Hamiltonian Action-wave Mechanics
(RHAM), all foundational problems of quantum mechanics are eliminated}, even
before we discuss the reconstruction of quantum mechanics by ensemble
averaging the action-wave dynamics. This is because the Schr{\"{o}}dinger
equation is not the single particle equation, whereas the foundational
problems mainly concern quantum effects related to the dynamical history and
observation of single physical systems. The new Hamiltonian action-wave
equation governs all dynamics, classical and quantum, and the solutions for
the foundational problems are to be sought there. Later, I will show that the
Schr{\"{o}}dinger equation is obtained as the ensemble average of Hamiltonian
action-wave dynamics and that the wavefunction in quantum mechanics is a
hybrid entity, with its real positive amplitude obtained as an ensemble
averaged probability density, combined with the complex action-wave.
Therefore, \emph{the wavefunction of quantum mechanics and the Schr\"{o}dinger
equation do not describe single particle dynamics (single dynamical history),
contrary to the present interpretations of quantum mechanics}. However, all
interference and correlations in quantum mechanics are correctly reproduced
from the local interference of the action-waves.

\section{Elimination of All Foundational Problems}

I have already discussed the universal nature of indeterminism and the absence
of a classical-quantum divide revealed by the new Hamiltonian action-wave
mechanics. The physical picture provided by the modified action-wave mechanics
is transparent. The particle, which is the carrier of dynamical quantities
like energy and momentum, has the dynamics linked to the temporal and spatial
derivatives of the action carried by associated action-waves. Action-waves
carry action $S(p_{i},x^{i})\propto\int p_{i}dx^{i}$, and not the dynamical
quantities $p_{i}$. In particular, action-waves do not possess energy or
momentum. The fundamental scale of action $\varepsilon$ limits the precision
to which the dynamical quantities can be specified in combination with the
coordinates. Particle or matter itself does not have wave property.
\emph{There are no matter waves}. What exists is a wave-particle connection,
as given by the Hamiltonian action-wave equation, and not wave-particle
duality. The long-held notion that matter behaves as waves, or as
matter-waves, was far from the actual physical fact, and it was a red herring.
Now I discuss each of the foundational problems and its definite solution
naturally arising from the single step of modifying the Hamiltonian action
mechanics to the equation $i\varepsilon\partial\zeta(S)/\partial t=H\zeta$. It
is significant that we do not need the Schr{\"{o}}dinger equation and the
wavefunction resulting from the ensemble average to discuss how the
foundational problems are eliminated, because they are not relevant for
analysing the dynamical history of single systems. The wavefunction of
standard QM turns out to be an ensemble averaged mathematical entity with no
physical counterpart for single dynamical history in space and time. Hence, I
refer to the `collapse of the wavefunction' as `collapse of the state' in
further discussion.

\subsection{Collapse of the State and Interference}

In RHAM, the dynamics of the particle contains significant scatter related to
the dephasing term in the dynamical equation when the experimental situation
involves multiple possibilities of dynamics comparable to the quantum of
action $\hbar$. We label these possibilities as $|\psi_{i}\rangle$,
anticipating convenience of notation later. (But, there is no distinction
between classical and quantum, for these states). As the material particle
encounters a fork of possibilities $|\psi_{i}\rangle$, it takes exactly one
$i=k$ with the probability determined by the experimental set up (beam
splitters, slits etc.) and the initial action-phase of the associated
action-wave at that instant. This initial phase is random and unknowable
within the quantum of action, and it has the intrinsic uncertainty $\hbar$.
This uncertainty translates to the choice of a random path. Since the material
particle has no wave property, it is never in superposition of physical
states. \emph{The particle is in one unique path (state) at any instant, in a
given event}. The action-wave splits as $a_{i}\exp(iS_{i}/\hbar)$ and
propagates in all paths or possibilities. The amplitude coefficients $a_{i}$
are determined by the experimental set up (like a beam splitter or an
aperture). The split action-waves obey the amplitude relation $\sum a_{i}%
^{2}=1$, as any other wave-like entity. Thus, for a symmetric beam splitter
$a_{i}=1/\sqrt{2}$, but this applies to only the action-waves and not for the
particle. The interference of action-waves determines the subsequent
probability of partition into other states (see calculation below). When a
measurement is done for any physical quantity, the result is the factual state
of the particle. \emph{When a measurement of position is done, the particle is
found where it actually is. }\emph{There is no collapse of the state}, because
the action-waves do not collapse and the particle is not in a superposition.
The action-waves, which do not carry energy or momentum, continue their
propagation, but it is irrelevant for further detail of the dynamics of the
particle. The relevant fact about the physical quantity `action' is that it is
an accumulated integral entity, linked to the space-time coordinate durations
in the dynamical history. Only the local relative action at the point of
overlap of multiple possible histories matters for interference and
probabilities, at the scale of the fundamental action of $\hbar\text{.}$ The
interactions of the action-waves involve exchange of only action. This is the
essence of wave-particle connection in RHAM. No conservation law is affected.

For repetitions of the particle dynamics with the state of the particle
prepared identically, the action-waves remain the same except for the
uncertainty $\hbar$ in action. This uncertainty results in the microscopic
stochasticity in the particle dynamics. Different histories in the ensemble
have different values for the physical observables. Now the ensemble average
defines the probabilities $\rho_{i}$ for the different $|\psi_{i}\rangle$. The
superposition like $a_{1}|\psi_{1}\rangle+a_{2}|\psi_{2}\rangle$ is only
applicable to the action-waves. For the particle, $\rho_{i}=A_{i}^{2}$ with
$\sum\rho_{i}=1$, where the real quantities $A_{i}$ refers to an ensemble of
dynamical histories of the particle in the state $|\psi_{i}\rangle$. It is
clear that the amplitudes $a_{i}$ of splitting of the action-waves at the
experimental set up are related to the ensemble averaged particle
probabilities as $\sqrt{\rho_{i}}=A_{i}=a_{i}$.

If the particle is allowed to propagate without interruption by detection to
the point where the action-waves are recombined, the phase difference after
interference determines the subsequent dynamics, stochastically, but only at
the tiny scale of $\varepsilon$.%
%TCIMACRO{\FRAME{ftbpFU}{4.6709in}{0.7603in}{0pt}{\Qcb{The factual quantum
%dynamics. A and B are two realizations of the spin-1/2 particle dynamics in a
%Stern-Gerlach device. The particle enters in a definite spin state and its
%dynamics continues in one of the two possibilities, Up or Down, in each run.
%There is no superposition of particle states. The action-waves corresponding
%to both the possibilities (dashed arrows) are present in both A and B. Their
%relative phase at the end, where paths are recombined, determines to
%probabilities of subsequent spin projections. }}{\Qlb{s-g}}{sg.png}%
%width 4.6709in;  height 0.7603in;  depth 0pt;  original-width 6.3444in;
%original-height 1.0083in;  cropleft "0";  croptop "1";  cropright "1";
%cropbottom "0";  filename 'SG.png';file-properties "XNPEU";}} }%
%BeginExpansion
\begin{figure}[ptb]%
\centering
\includegraphics[
natheight=1.008300in,
natwidth=6.344400in,
height=0.7603in,
width=4.6709in
]%
{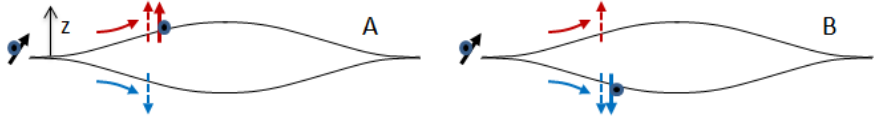}%
\caption{The factual quantum dynamics. A and B are two realizations of the
spin-1/2 particle dynamics in a Stern-Gerlach device. The particle enters in a
definite spin state and its dynamics continues in one of the two
possibilities, Up or Down, in each run. There is no superposition of particle
states. The action-waves corresponding to both the possibilities (dashed
arrows) are present in both A and B. Their relative phase at the end, where
paths are recombined, determines to probabilities of subsequent spin
projections. }%
\label{s-g}%
\end{figure}
%EndExpansion

Consider a generic symmetric interferometer for neutrons, figure \ref{s-g},
with the field $B$ and its gradient $B^{\prime}$ in the z-direction. At the
point of entry in the apparatus, the random initial phase of x-polarized
neutrons results in the random choice of $|z+\rangle$ or $|z-\rangle$ state
and the path a particular neutron actually takes. The particle propagates
exactly in one of the paths and unique state ($|z+\rangle$ or $|z-\rangle$) in
each event, with no superposition. Measurement before path closure results in
the factual state of the particle; there is no collapse. The action-waves
corresponding to both states are present in every event, accumulating the
dynamical phase $\int pdx$, and the phase from the interaction $V(x)/\hbar
=\mu\cdot B/\hbar$. If there are spin flippers in the paths, \textit{the
actual spin flip with the energy exchange } $\Delta E=2\mu_{n}B$
\textit{happens only for the particle, where it is actually propagating}. The
action-wave changes its phase by $\pi/2$ as it passes the spin flipper, but
there is no exchange of energy or `flip of spin'.

The initial action-waves corresponding to the possibilities of the particle
states $|x+\rangle$ or $x-\rangle$ would split at the entrance port of a
Stern-Gerlach interferometer as
\begin{align}
|x+\rangle &  =\frac{1}{\sqrt{2}}|z+\rangle+\frac{1}{\sqrt{2}}|z-\rangle
\nonumber\\
|x-\rangle &  =\frac{1}{\sqrt{2}}|z+\rangle-\frac{1}{\sqrt{2}}|z-\rangle
\end{align}
This is very similar to a Mach-Zehnder interferometer with a polarizing
beam-splitter and two orthogonal input states. However, we stress that the
particle enters in only one of the path-states, $|z+\rangle$ or $|z-\rangle$.

The time evolved action-waves superpose at the exit as%

\begin{equation}
|x^{\prime}+\rangle=\frac{1}{\sqrt{2}}|z+\rangle e^{i\omega_{+}t}+\frac
{1}{\sqrt{2}}|z-\rangle e^{i\omega_{-}t}%
\end{equation}
The quantity $\omega_{\pm}\equiv\pm\mu_{n}B/\hbar$. Note that the probability
for the particles to be in the $|z+\rangle$ or the $|z-\rangle$ does not
change due to the phase evolution, before interference at exit. Now the
probability for getting the particle in the state $|x-\rangle$ can be calculated,%

\begin{equation}
P(x + \rightarrow x -) =\vert\langle x -\vert x^{ \prime} + \rangle\vert^{2}
=\left\vert \frac{1}{2} \left(  e^{i \omega_{ +} t} -e^{i \omega_{ -}
t}\right)  \right\vert ^{2} =\frac{1}{2} \left(  1 -\cos\frac{2 \mu_{n}
B}{\hbar} t\right)
\end{equation}

\emph{We have reproduced the most important feature of single particle quantum
interference, while avoiding the collapse of the quantum state}. This is the
feature that Feynman described as the `only mystery' in quantum mechanics.
When the particle emerges out, the resultant phase of the action-waves from
both the paths determines the subsequent probability to be found in states
$|x\pm\rangle\text{,}$ $|y\pm\rangle$ etc. The probability for the other
possibility is $P(x+\rightarrow x+)=1-P(x+\rightarrow x-)$.

As other examples, a massive neutrino of one flavour is in \emph{one and only
one of the three mass states and never in superposition of the three}. The
action-waves corresponding to the mass states superpose and interfere,
resulting in oscillations in the probabilities for different flavour states.
An electron neutrino can be in only one of the three mass states at origin,
and remains so throughout its propagation. The interference of the
action-waves changes only the probabilities for flavour states. (The details
are worked out in the appendix 1, for the case of two states). This example is
important in the context of the theory of gravity and quantum gravity, because
RHAM makes clear that there is no superposition of different masses, energies,
and long-range gravitational fields. Similarly, an atom is never in a
superposition of two positions or two (or more) energy states; only the
action-waves are, which carry no energy.

The reconstruction generalizes to the relativistic situation and the Dirac
equation $i\hbar\dot{\psi}=\hat{H}_{D}\psi$, with the same Hamiltonian
action-wave equation as the basis for relativistic particle dynamics as well.
In fact, the fundamental requirement that the time evolution is of first order
arises from the parent action-wave equation. Later I will show how this
naturally incorporates the time evolution of probability and the Born's rule,
independent of the action-wave equation of dynamics. This resolves the
foundational issue of `zitterbewegung', identifying it as the ensemble
average, and not as a feature of single particle dynamics (see section 4.2 and
appendix 2).%
%TCIMACRO{\FRAME{ftbpFU}{4.0058in}{1.1204in}{0pt}{\Qcb{The separable particle
%states in the situation that is conventionally treated as entangled and
%non-separable. The particles have definite physical states in each dynamical
%history. Two pairs of correlated action-waves are operative in each history
%and their local interferences correctly give the quantum correlation without
%nonlocality (section 3.4).}}{\Qlb{separable}}{separable.png}%
%width 4.0058in;  height 1.1204in;  depth 0pt;  original-width 5.8687in;
%original-height 1.6201in;  cropleft "0";  croptop "1";  cropright "1";
%cropbottom "0";  filename 'separable.png';file-properties "XNPEU";}} }%
%BeginExpansion
\begin{figure}[ptb]%
\centering
\includegraphics[
natheight=1.620100in,
natwidth=5.868700in,
height=1.1204in,
width=4.0058in
]%
{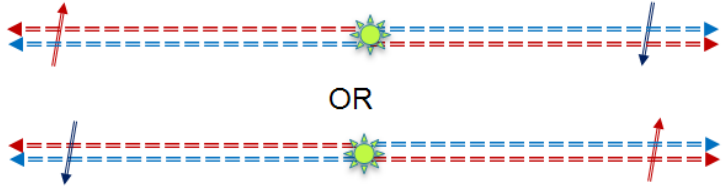}%
\caption{The separable particle states in the situation that is conventionally
treated as entangled and non-separable. The particles have definite physical
states in each dynamical history. Two pairs of correlated action-waves are
operative in each history and their local interferences correctly give the
quantum correlation without nonlocality (section 3.4).}%
\label{separable}%
\end{figure}
%EndExpansion

A multi-particle `entangled' system in RHAM is equally simple. The
superposition represented as $c_{1}|+\rangle_{1}|-\rangle_{2}+c_{2}%
|-\rangle_{1}|+\rangle_{2}$, and conventionally interpreted as the
non-separable entangled state, is applicable only to the associated
action-waves (fig. \ref{separable}). \ The joint physical state of the
\emph{two particles} is always a unique unentangled separable state, like
$|+\rangle_{1}|-\rangle_{2}$ or $|-\rangle_{1}|+\rangle_{2}$. The single
dynamical history of a pair of particles states is not entangled. The
probabilities for measurement results in a general basis are determined by the
\emph{local relative phase} of the interference of the \emph{correlated pairs
of action-waves} (see section 3.4). Hence there is no collapse of the state.
\emph{Nor is any nonlocality}. Ensemble average gives $\left\vert
c_{1}\right\vert $ and $\left\vert c_{2}\right\vert $ for the average particle
states and determines the hybrid wavefunction; \emph{`entanglement' of
particle states is apparent and an ensemble statement}. RHAM and the ensemble
averaged RQM deconstruct the enigma of quantum entanglement, restoring the
physicality and separability of the particle states. I will show explicitly
(section 3.4) that the action-waves interfere locally for each particle of the
pair to give all correlations correctly, after addressing the vital quantum
measurement problem.

\subsection{Quantum Measurement Problem}

A quantum system that can be in the superposition $\sum c_{i}|\psi_{i}\rangle$
is conventionally thought to be in the correlated entangled state after
interaction with an apparatus with pointer states $|P_{j}\rangle$,
\begin{equation}
|\psi\rangle_{sA}=\sum c_{k}|\psi_{k}\rangle|P_{k}\rangle\label{sys-App}%
\end{equation}
However, we have already seen in the example of single particles or a pair of
particles that only the action-waves corresponding to the different possible
physical states are in superposition, and that the matter states are unique
and distinct at every event. \emph{Entanglement between the system and
apparatus, or between any two systems, is an ensemble apparition}. The
superposition in equation (\ref{sys-App}) is applicable only to the action
waves; for particle states it is in fact an ensemble averaged expression. In
each trial, a particular state $|\psi_{n}\rangle$ of the system, determined by
the local interference of the action-waves, results in a unique correlated
pointer state $|P_{n}\rangle$ of the apparatus after the interaction, giving
the joint product state $|\psi_{n}\rangle|P_{n}\rangle$. \emph{The joint state
after interaction at any measurement event is exactly one of the possible}
$|\psi_{k}\rangle|P_{k}\rangle$. Since there is no collapse of the state of
either the system or the apparatus, the quantum measurement problem is solved
completely. The joint state $|\psi_{k}\rangle|P_{k}\rangle$ changes from trial
to trial. The ensemble average of such multiple measurements has the relative
probability $\left\vert c_{k}\right\vert ^{2}$. (It is this set of
probabilities that decoherence models reproduce, and not what happens in
single measurement events). The macroscopic apparatus is characterised by the
stable action-phase of its state, because the quantum uncertainty in the phase
amounting to a few $\hbar$ is insignificant compared to the gross action of
the large physical system. Otherwise, there is no difference between
macroscopic and microscopic matter, in RHAM and RQM.

\subsection{The Quantum-Classical Divide}

The starting point of the reconstruction was the generalized Hamiltonian
action-wave mechanics, with universal scope of all dynamics. Quantum mechanics
is just the ensemble average of universal mechanics. Therefore, there is no
classical-quantum divide in particle mechanics. For a single particle or
dynamical history, universal mechanics of Hamiltonian action-wave equation is
applicable. Averaged over the ensemble of histories, we get the Schr{\"{o}}%
dinger equation and quantum mechanics (see section 4). The action-wave
equation is more fundamental. There is no change or transition of the
dynamical law when we pass from the macroscopic world to the microscopic
world. Neither is any involvement of the observer and `consciousness', beyond
what is obvious and familiar in the macroscopic world. We already discussed
how the perceived entanglement in measurement pertains to the ensemble and not
to the individual act of observation. There is no chain of observation
entanglements, to be eventually resolved by a conscious observer. All
speculations over the role of the observer are reduced to the microscopic
disturbances to the action-waves, involving a few quanta of action. The
quantum phase uncontrollably changes during the exchange of a quantum of
action. This explains all cases of inherent indeterminism during measurement
and the effects on quantum interference, without invoking direct perturbations
on the dynamical content like the momentum of the particle \cite{Unni-PRA}.
The action-waves of the macroscopic systems interfere over tiny
spatio-temporal intervals, much smaller than atomic scales, and thus quantum
interference is not effective in determining their dynamics. Hence, for
macroscopic objects the Hamiltonian action wave equation $i\varepsilon
\frac{\partial\zeta}{\partial t}=H\zeta$ can be reduced to the approximate
Hamilton-Jacobi dynamical equation $\frac{\partial S}{\partial t}=-H$.

\subsection{Entanglement and Correlations}

We conclude the discussion of the elimination of foundational problems with
the \emph{solution of the hard problem of the correlations of the two-particle
entangled system} in which the enigma of quantum mechanics appears in the most
pronounced way. It is this problem that gave rise to most debates on quantum
mechanics, initiated by the Einstein-Podolsky-Rosen incompleteness argument
that established the incompatibility between Einstein locality and the
wavefunction description of physical states in QM (see appendix 4). As well
known, two-particle correlations were discussed also in the context of certain
classical statistical theories with `hidden' statistical variables, which were
advanced to replace the theory of quantum mechanics, with an upper limit on
such correlations represented by the Bell's inequalities \cite{Bell}. I have
shown elsewhere that such local hidden variable theories are incompatible with
the fundamental conservation laws, and hence unphysical
\cite{Unni-EPL,Unni-Pramana}. In any case, all debates on multi-particle
correlations can rest now, in the light of the factual physical situation
revealed by RHAM.

\begin{figure}[ptb]%
\centering
\includegraphics[
natheight=1.309800in,
natwidth=10.136800in,
height=0.7389in,
width=5.6161in
]%
{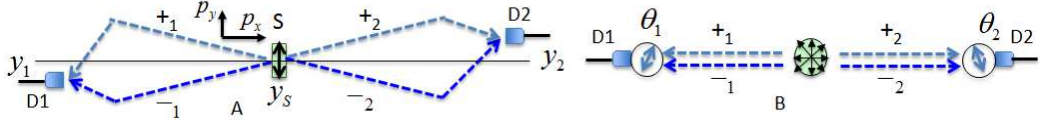}%
\caption{Quantum correlations derived from the local and independent
interference of the action-waves at the detectors. (The definite physical
states of the particles are not shown). The uncertainty in the initial phase
at the source washes out interference at individual detectors, but their
correlation is a definite function with 100\% `visibility'. For identical
settings, we get perfect correlation without any nonlocality. }%
\label{q-corr}%
\end{figure}

Consider a two-particle state that is correlated in momentum, but restricted
to two possible values $\pm$ (two paths) in an interference experiment, as
shown in Fig. \ref{q-corr}A. The particles are always in some definite
correlated physical state, either $|+\rangle_{1}|-\rangle_{2}$ or
$|-\rangle_{1}|+\rangle_{2}$, in a given pair-event. \ There is no
superposition of two-particle states. Two \emph{pairs of action-waves,} (
$+_{1},-_{1}$ ) and ( $+_{2},-_{2}$ ), associated with the two particles are
present in each event. The action-waves are correlated at the source (or
interaction point) through the conservation
laws~\cite{Unni-EPL,Unni-Pramana,Unni-FPL}. When the dynamical quantity
$p_{s}$ is conserved, it reflects in the phase of the action-waves as
$\exp(ip_{s}x^{s})$, with $p_{s}=p_{1}+p_{2}$. \emph{The sum of the initial
phases are fixed by the conservation constraint, but the individual phases are
random}. \ We will see that this directly translates to the feature that
individual local measurements on each particle give random results. But their
correlation visibility can be 100\%, depending on the degree to which the
conservation constraint $p_{s}=p_{1}+p_{2}$ is maintained during the
generation and propagation of the particles from the source.

I will now derive the two-particle correlation from entirely local
interference of action-waves, retaining the independent physical states of the
particles, eliminating nonseparability and nonlocality, and answering the EPR
query.%
%TCIMACRO{\FRAME{ftbpFU}{5.6161in}{0.7389in}{0pt}{\Qcb{Quantum correlations
%derived from the local and independent interference of the action-waves at the
%detectors. (The definite physical states of the particles are not shown). The
%uncertainty in the initial phase at the source washes out interference at
%individual detectors, but their correlation is a definite function with 100\%
%`visibility'. For identical settings, we get perfect correlation without any
%nonlocality. }}{\Qlb{q-corr}}{qcorrelation.png}%
%width 5.6161in;  height 0.7389in;  depth 0pt;  original-width 10.1368in;
%original-height 1.3098in;  cropleft "0";  croptop "1";  cropright "1";
%cropbottom "0";  filename 'QCorrelation.png';file-properties "XNPEU";}} }%
%BeginExpansion
%EndExpansion

Referring to Fig. \ref{q-corr}A, the source emits the particles with opposite
momenta $\left(  p_{+1},p_{-2}\right)  $ or $\left(  p_{-1},p_{+2}\right)  $.
An important requirement on the source of a correlated pair of particles is
that the physical nature of the source should not limit the correlation. In
the case of correlation in momenta, the size of the source should be larger
than $h/\delta p$ for getting correlated particles, where $\delta p$ signifies
the degree of the lack of correlation (this requirement is present both in
RHAM and conventional quantum mechanics). Thus, the point of origin $y_{s}$ of
the particles varies by more than $h/\Delta p_{\pm}$ event to event,
stochastically. Now, two pairs of correlated action-waves from $y_{s}$ are
present in each event, one pair for each particle; $p_{+1}$ and $p_{-1}$
associated with particle \#1 and $p_{+2}$ and $p_{+2}$ with particle \#2. They
interfere locally at the detector D1 at $y_{1}$ and at the detector D2 at
$y_{2}$.%

\begin{align}
\vert1 ,y_{1} \rangle=\frac{1}{\sqrt{2}} e^{i p_{ +1} \left(  y_{1}
-y_{s}\right)  /\hbar} +\frac{1}{\sqrt{2}} e^{i p_{ -1} \left(  y_{1}
-y_{s}\right)  /\hbar}\nonumber\\
\vert2 ,y_{2} \rangle=\frac{1}{\sqrt{2}} e^{i p_{ -2} \left(  y_{2}
-y_{s}\right)  /\hbar} +\frac{1}{\sqrt{2}} e^{i p_{ +2} \left(  y_{2}
-y_{s}\right)  /\hbar}%
\end{align}

The action-wave $p_{+1}$ is always coupled with the $p_{-2}$ wave and the
$p_{-1}$ wave with $p_{+2}$ wave, since they are correlated; there are no
$\left(  p_{+1},p_{+2}\right)  $ \ or $\left(  p_{-1},p_{-2}\right)  $
combinations. Because of the conservation constraint, the $p_{+1},p_{-2}$
action-waves are like a single coherent phase-wave; so are $p_{-1},p_{+2}$.
Also, the correlation requires that the transverse momenta obey $p_{+}=-p_{-}
$. The detected intensity at the detector D1 of area $dS$ is $\left\vert
|1,y_{1}\rangle\right\vert ^{2}dS$.%

\begin{align}
\left\vert \vert1 ,y_{1} \rangle\right\vert ^{2}  &  =1 +\frac{1}{2} e^{i p_{
+1} \left(  y_{1} -y_{s}\right)  /\hbar} e^{ -i p_{ -1} \left(  y_{1}
-y_{s}\right)  /\hbar} +\frac{1}{2} e^{i p_{ -1} \left(  y_{1} -y_{s}\right)
/\hbar} e^{ -i p_{ +1} \left(  y_{1} -y_{s}\right)  /\hbar}\nonumber\\
&  =1 +\frac{1}{2} e^{i \left(  p_{ +} -p_{ -}\right)  \left(  y_{1}
-y_{s}\right)  /\hbar} +\frac{1}{2} e^{ -i \left(  p_{ +} -p_{ -}\right)
\left(  y_{1} -y_{s}\right)  /\hbar} =1 +\cos\left(  \Delta k \left(  y_{1}
-y_{s}\right)  \right)
\end{align}

We have written $p/\hbar=k$. For particle \#2, we get a similar expression
$\left\vert |2,y_{2}\rangle\right\vert ^{2}=1+\cos\left(  \Delta k\left(
y_{2}+y_{s}\right)  \right)  $. Though these are the familiar cosine forms of
interference, \emph{there is no single particle interference } because the
source point $y_{s}$ is a stochastic quantity in the extended source, and
$y_{s}\Delta k\geq2\pi$. Thus, the average intensity at any detector position
is $\langle\left\vert |1,y_{1}\rangle\right\vert ^{2}\rangle=\langle\left\vert
|2,y_{2}\rangle\right\vert ^{2}\rangle=1$, a uniform intensity.

However, for the two-particle coincidence detection in the two detectors
involves simultaneous, but independent, local interference of the action-waves
at location 1 and 2, as a simple product of the two amplitudes,
\begin{equation}
|1,y_{1}\rangle|2,y_{2}\rangle=\left(  \frac{1}{\sqrt{2}}e^{ik_{+1}\left(
y_{1}-y_{s}\right)  }+\frac{1}{\sqrt{2}}e^{ik_{-1}\left(  y_{1}-y_{s}\right)
}\right)  _{1}\left(  \frac{1}{\sqrt{2}}e^{ik_{-2}\left(  y_{2}-y_{s}\right)
}+\frac{1}{\sqrt{2}}e^{ik_{+2}\left(  y_{2}-y_{s}\right)  }\right)  _{2}%
\end{equation}
Keeping in mind that there is no $\left(  k_{+1},k_{+2}\right)  $ \ or
$\left(  k_{-1},k_{-2}\right)  $ combinations, \ we get
\begin{equation}
|1,y_{1}\rangle|2,y_{2}\rangle=\frac{1}{\sqrt{2}}e^{ik_{+1}\left(  y_{1}%
-y_{s}\right)  }\frac{1}{\sqrt{2}}e^{ik_{-2}\left(  y_{2}-y_{s}\right)
}+\frac{1}{\sqrt{2}}e^{ik_{-1}\left(  y_{1}-y_{s}\right)  }\frac{1}{\sqrt{2}%
}e^{ik_{+2}\left(  y_{2}-y_{s}\right)  } \label{Entanglement}%
\end{equation}
This equation holds the important physical point. \emph{There is no
`entanglement' in the physical states of each pair of particles}. Entanglement
is just the correlation of the action-waves, explicit in equation
(\ref{Entanglement}). \emph{Particle states are separable}.

The detected intensity-intensity correlation is%

\begin{align}
\left\vert |1,y_{1}\rangle|2,y_{2}\rangle\right\vert ^{2}  &  =\frac{1}%
{2}+\frac{1}{4}e^{i\left(  k_{-}-k_{+}\right)  \left(  y_{1}-y_{s}\right)
}e^{-i\left(  k_{-}-k_{+}\right)  \left(  y_{2}-y_{s}\right)  }+\frac{1}%
{4}e^{-i\left(  k_{-}-k_{+}\right)  \left(  y_{1}-y_{s}\right)  }e^{i\left(
k_{-}-k_{+}\right)  \left(  y_{2}-y_{s}\right)  }\nonumber\\
&  =\frac{1}{2}+\frac{1}{4}e^{i\Delta k_{y}\left(  y_{1}-y_{2}\right)  }%
+\frac{1}{4}e^{-i\Delta k_{y}\left(  y_{1}-y_{2}\right)  }=\frac{1}{2}\left(
1+\cos\left(  \Delta k_{y}\left(  y_{1}-y_{2}\right)  \right)  \right)
\end{align}
\emph{This is an interference pattern with 100\% visibility}, visible by
fixing one detector and scanning the other. There is no collapse of the state
and there is no nonlocality. A measurement at D1 has no influence whatsoever
over what is detected at D2. The two are independent local interferences and
respect strict Einstein locality. However, when correlated with time stamps,
perfect interferometric correlation results. Any mixture or lack of
correlation in the action-waves results in the reduction of visibility,
eventually dropping to 50\% for the thermal state, as in the Hanburry
Brown-Twiss correlation.

The correlation of the particles of a spin singlet that epitomises all the
quantum mysteries is treated similarly. The only difference is that the
correlation is in angular momentum (spin). When the pair of particles emerge
from the source with total spin zero, \emph{they are randomly but entirely in
only one of the two possible joint states} $|S_{\pm}\rangle=|+\rangle
_{1}|-\rangle_{2}$ or $|S_{\mp}\rangle=|-\rangle_{1}|+\rangle_{2}$, with no
superposition. The action-waves corresponding to both states are present in
every event (similar to two anti-correlated helicity states). The entangled
state in conventional QM is the ensemble averaged hybrid wavefunction
$\langle|S=0\rangle\rangle=\left(  |+\rangle_{1}|-\rangle_{2}-|-\rangle
_{1}|+\rangle_{2}\right)  /\sqrt{2}$ in any basis, as explained in the next
section. \emph{This entanglement is not a statement on each pair, but over the
ensemble}.

For the case of spins, the initial random angular orientation $\varphi$ in
fig. \ref{q-corr}B is like the random origin $y_{s}$ in fig. \ref{q-corr}A.
The spin analyzer orientation $\theta_{i}$ corresponds to a stable phase,
exactly as the detector position in the example of the interference.
Action-waves associated with each particle superpose and interfere, locally at
analysers 1 and 2. Instead of the factor $k_{+1}\left(  y_{1}-y_{s}\right)  $
etc. in the two-path interference case earlier, here we have $s_{+}\left(
\theta_{i}-\varphi\right)  $ and $s_{-}\left(  \theta_{i}-\varphi\right)  $ in
the local interferences at analysers \#1 and \#2.%

\begin{align}
\vert1 ,\theta_{1} \rangle=\frac{1}{\sqrt{2}}\vert+ \rangle_{1} e^{i s_{ +}
\theta_{1}} e^{ -i s_{ +} \varphi} +\frac{1}{\sqrt{2}}\vert- \rangle_{1} e^{i
s_{ -} \theta_{1}} e^{ -i s_{ -} \varphi}\nonumber\\
\vert2 ,\theta_{2} \rangle=\frac{1}{\sqrt{2}}\vert+ \rangle_{2} e^{i s_{ +}
\theta_{2}} e^{ -i s_{ +} \varphi} +\frac{1}{\sqrt{2}}\vert- \rangle_{2} e^{i
s_{ -} \theta_{2}} e^{ -i s_{ -} \varphi}%
\end{align}

For independent detection, say particle \#1 in the state $|+\rangle$, with
spin analyzer oriented at $\theta_{1}$ the situation is analogous to
interferometer in fig. \ref{s-g}, but now with a random phase $\varphi$. We
get the probability%

\begin{align}
P\left(  1+,\theta_{1}\right)   &  =\left\vert \langle1+||1,\theta_{1}%
\rangle\right\vert ^{2}=\left\vert \frac{1}{2}e^{is_{+}\theta_{1}}%
e^{-is_{+}\varphi}+\frac{1}{2}e^{is_{-}\theta_{1}}e^{-is_{-}\varphi
}\right\vert ^{2}\nonumber\\
&  =\frac{1}{2}+\frac{1}{2}\cos\left(  \Delta s\left(  \theta_{1}%
-\varphi\right)  \right)
\end{align}
The probability $P(1-,\theta_{1})=1-P(1+,\theta_{1})$. Since $\varphi$ is
stochastic in the interval $\left(  0,2\pi\right)  $, the average of
$\cos\Delta s\left(  \varphi-\theta_{1}\right)  $ is zero and there is
\emph{uniform probability } $1/2$ \emph{for detecting} $+1$ and $-1$ \emph{for
the spin projections, for any setting of the apparatus}.

The joint detection is with the constraint that the $|+\rangle_{1}$
action-wave is correlated with the $|-\rangle_{2}$ wave and the $|-\rangle
_{1}$ action-wave is with the $|+\rangle_{2}$ wave. Analogous to a
double-interferometer, as in fig. \ref{q-corr}A, there are no $|+\rangle
_{1}|+\rangle_{2}$ wave or $|-\rangle_{1}|-\rangle_{2}$ combinations.
\begin{equation}
|1,2\rangle=\frac{1}{2}|+\rangle_{1}e^{is_{+}\theta_{1}}e^{-is_{+}\varphi
}|-\rangle_{2}e^{is_{-}\theta_{2}}e^{-is_{-}\varphi}+\frac{1}{2}|-\rangle
_{1}e^{is_{-}\theta_{1}}e^{-is_{-}\varphi}|+\rangle_{2}e^{is_{+}\theta_{2}%
}e^{-is_{+}\varphi}%
\end{equation}
The probability for joint measurement of $(+1,-1)$ or $(-1,+1)$ for the spin
projections with their product $-1$ is
\begin{align}
P_{+-,-+}  &  =\left\vert |1_{+}2_{-},1_{-}2_{+}\rangle\right\vert ^{2}%
=\frac{1}{2}+\frac{1}{4}e^{-i\left(  s_{+}-s_{-}\right)  \left(  \theta
_{1}-\theta_{2}\right)  }+\frac{1}{4}e^{i\left(  s_{+}-s_{-}\right)  \left(
\theta_{1}-\theta_{2}\right)  }\nonumber\\
&  =\frac{1}{2}\left(  1+\cos(\Delta s\left(  \theta_{1}-\theta_{2}\right)
)\right)  =\cos^{2}\left(  \theta_{1}-\theta_{2}\right)  /2
\end{align}
Then the probability to get $(+1,+1)$ or $(-1,-1)$ for the spin projections
with their product $+1$ is
\begin{equation}
P_{++,--}=\left\vert |1_{+}2_{+},1_{-}2_{-}\rangle\right\vert ^{2}%
\rangle=1-\cos^{2}\left(  \theta_{1}-\theta_{2}\right)  /2=\sin^{2}\left(
\theta_{1}-\theta_{2}\right)  /2
\end{equation}
The final spin correlation function for particles from a spin singlet
$\left\vert S=0\right\rangle $ is
\begin{equation}
C(\theta_{1}-\theta_{2})=-1\times P_{+-,-+}+1\times P_{++,--}=-\cos\left(
\theta_{1}-\theta_{2}\right)
\end{equation}
I have derived the singlet correlation from the local interference of the
action waves of RHAM, without nonlocality and collapse of the states. The
derivation did not even refer to the conventional entangled wavefunction. When
$\left(  \theta_{1}-\theta_{2}\right)  =0$, \emph{we get perfect
anti-correlation event by event for every pair of particles, irrespective of
the individual setting of the analyzers or the distance between the
particles}. Resolution of this problem, considered the greatest quantum
mystery, shows the correctness and scope of the action-wave mechanics and the
reconstructed quantum mechanics. All familiar results involving bipartite and
multi-particle correlations, like teleportation, happen through the
action-wave correlation, while particle states remain separable and
unentangled. There are no exceptions.

\section{Reconstruction of Quantum Mechanics}

\subsection{General Description}

Deepening dynamics by modifying the Hamiltonian action mechanics with the new
wave equation $i\hbar\partial\zeta(x,t)/\partial t=H\zeta$ enabled the
complete elimination of all the foundational problems that plagued quantum
mechanics. It remains to show that the Schr\"{o}dinger equation and quantum
mechanics emerge naturally and entirely from the ensemble average of the
dynamical histories of Hamiltonian action-wave mechanics. There is no
difference in the dynamical equation between macroscopic and microscopic
dynamics. There is a fundamental uncertainty in all dynamics, because the
action is specified only up to the fundamental scale, with the uncertainty
$\Delta\int p_{i}dx^{i}=\hbar$. Therefore, the equation $\nabla S(x,t)=p$ also
is an approximation, ignoring the fundamental uncertainty due to the quantum
of action $\hbar$. \emph{Classical mechanics, or mechanics in its core form,
is not deterministic, contrary to the generally held view over centuries}. The
momentum of the particle is related to the action-wave, $p\zeta=-i\varepsilon
\nabla\zeta$. Therefore, the dynamics of one particle differs from another
identical one by such a difference in the action. This irreducible
stochasticity manifests significantly only at microscopic scale, in the
dynamics of fundamental particles, atoms and light.

When we come to microscopic dynamics, the dephasing term in the time evolution
of action, $\partial S/\partial t=i\hbar\nabla^{2}S/2m$, becomes significant
and the dynamical histories vary considerably from particle to particle. Then,
\emph{we have no choice, but to describe dynamics in a statistical framework,
ensemble averaged}. Averaging over the particle dynamics leads to probability
densities $\rho_{i}$ for different physical states $\left\vert s_{i}%
\right\rangle .$ Since the action-wave amplitudes in different possibilities
obey $a_{i}^{2}=\rho_{i}$, the ensemble average of the action-wave equation
has the same basic form, $i\hbar\partial f(x,t)/\partial t=Hf(x,t)$, but now
with a new ensemble averaged function $f(x,t)=\sqrt{\rho(x,t)}\left\langle
\zeta(x,t)\right\rangle $. The Schr\"{o}dinger equation describes the time
evolution of the hybrid ensemble function $f(x,t)$.

The discussion of foundational problems in quantum mechanics in the past were
all focussed on the analysis of the Schr{\"{o}}dinger equation, because it was
assumed that it was the microscopic fundamental equation from which
macroscopic mechanics would emerge as an effective theory. This expectation
was not realised, and the reason is clear now. The Hamiltonian action-wave
equation is the fundamental equation for all dynamics and the Schr{\"{o}}%
dinger equation is the result of its ensemble averaging.

\subsection{Details of Reconstruction}

Once we are in the domain of statistical description of ensemble averaged
dynamics, which is the only way to describe dynamics in the microscopic world,
we need to consider the time evolution of the probability density explicitly,
along with the time evolution of dynamical quantities. I will show now that
the ensemble equation in which both are integrated is the Schr\"{o}dinger equation.

The classical conservation (continuity) equation for the probability density
is
\begin{equation}
\frac{\partial\rho(x,t)}{\partial t}=-\nabla\cdot(\rho v)=-\rho\nabla\cdot
v-v\cdot\nabla\rho\label{Continuity}%
\end{equation}
Thus, dynamics enters the continuity equation through the particle velocities.
With the modified action-wave equation, \emph{this continuity equation is not
complete} because of the dephasing term ($i\hbar\nabla^{2}S/2m$) in the
action-wave equation contributes an additional microscopic term related to
$\nabla^{2}S=\nabla\cdot v$. We have already seen how dephasing causes a
dissipationless `diffusion' of the amplitude of the ensemble of action-waves
(refer to figure \ref{dephasing}). Therefore, we get the corresponding
\emph{diffusion equation for the amplitude of the action-waves, with the same
imaginary diffusion constant},
\begin{equation}
\frac{\partial a(x,t)}{\partial t}=\frac{i\hbar}{2m}\nabla^{2}a \label{diff-a}%
\end{equation}
Now we note the important connection between the amplitude $a_{i}$ of the
action-waves and the ensemble averaged probabilities of particle dynamics,
$\sqrt{\rho_{i}}=A_{i}=a_{i}$, discussed in section 3.1.

Since the positive real $\rho(x,t)=A^{2}(x,t)$, the continuity equation
(\ref{Continuity}) for $A(x,t)$ is completed by adding the term in equation
(\ref{diff-a})
\begin{equation}
\frac{\partial A(x,t)}{\partial t}=-\frac{A}{2}\nabla\cdot v-v\cdot\nabla
A\rightarrow-\frac{A}{2m}\nabla^{2}S-\frac{1}{m}\nabla S\cdot\nabla
A+\frac{i\hbar\chi}{2mA}\nabla^{2}A
\end{equation}

We can recast this \emph{equation of constraint} as a `wave equation' by
defining the pseudo-wave $\chi(x,t)=\langle\rho^{1/2}\rangle\exp(i\phi)\equiv
A\exp(i\phi)$. \ The mathematical independence of $\rho$ from the dummy phase
$\phi$ is ensured with $\rho=A^{2}=\chi\chi^{\ast}$. \textit{This is the exact
Born's rule}. We stress that it is universally applicable, for dynamics at all scales.

The appropriate equation for the function $\chi(x,t)$ that contains all the
terms in the equation for $\partial A/\partial t$ is
\begin{equation}
\partial\chi/\partial t=i\chi\frac{\partial\phi}{\partial t}+\frac{\chi}%
{A}\frac{\partial A}{\partial t}=i\chi\frac{\partial\phi}{\partial t}%
-\frac{\chi}{2m}\nabla^{2}S-\frac{\chi}{Am}\nabla S\cdot\nabla A+\frac
{i\hbar\chi}{2mA}\nabla^{2}A
\end{equation}
The time evolution $\partial A/\partial t$ is represented by the last three
terms in the evolution equation for $\partial\chi/\partial t$. \ The first
term concerns the evolution of the dummy phase of the complex function
representing $\sqrt{\rho}$. Up to this point, the discussion \ was general and
no aspect of dynamics was specifically involved, except expressing the
velocity in term of the gradient of the action-wave in the ensemble continuity
equation. In particular, $\phi$ has no connection to dynamics or probability.
Now we take the step that takes us directly to the theory of quantum mechanics.

The term $\partial\phi/\partial t$ in $\partial\chi/\partial t\text{{}}$ is
redundant as it stands because $\phi$ is a dummy phase with no relation to
dynamics. It is striking that the ensemble of particle dynamics and the
continuity of its probability density can be combined if we elevate the dummy
quantity $\phi$ to be the action $S(x,t)/\hbar$ in a \emph{hybrid function}
$\psi(x,t)=A(x,t)\zeta(x,t)\text{. }$Both $A(x,t)$ and $\zeta(x,t)$ are
averaged over the ensemble of dynamical histories. For dynamics starting from
the same physical state, the action-waves remain the same, but the particle
dynamics is stochastic, as I discussed in section 3.

$\text{ The new hybrid `wavefunction' is }\text{{}}\psi(x,t)=A(x,t)\exp
(iS(x,t)/\hbar)$. Then
\begin{equation}
i\hbar\partial\psi/\partial t=-\psi\frac{\partial S}{\partial t}+\frac
{i\hbar\psi}{A}\frac{\partial A}{\partial t}=-\psi\frac{\partial S}{\partial
t}-\frac{i\hbar\psi}{2m}\nabla^{2}S-\frac{i\hbar\psi}{Am}\nabla S\cdot\nabla
A-\frac{\hbar^{2}\psi}{2mA}\nabla^{2}A \label{lhs-sch}%
\end{equation}
Here, $\bar{A}\equiv A=\sqrt{\rho}$ is an ensemble averaged quantity and
$S(x,t)/\hbar$ is the phase of the action-wave of single particle dynamics. We
see clearly how the ensemble average of the time evolution equation of the
action-wave over dynamical histories naturally leads to the time evolution
equation for the amplitude of the action-waves, and through that the time
evolution of probabilities.

Thus, we fuse the action-wave and the ensemble averaged probability density
(its square root) into a new abstract wave-function, representing the ensemble
average of many dynamical histories with the same action-waves. This has no
existence in space and time due to the fact that the real amplitude $\bar
{A}(x,t)$ is an ensemble averaged quantity. Thus, the \emph{wavefunction is
not the action-wave. Nor is it a probability wave}. Since there are only
action-waves in reality and not matter-waves, wavefunction is not a
matter-wave either. \emph{The wavefunction }$\psi(x,t)$ \emph{an abstract
ensemble entity with no direct ontological significance}. Its mathematical
properties are determined by the way it is synthesised; it is complex valued
(action-wave) and square integrable (finite probabilities).

It is remarkable that the fundamental Hamiltonian action-wave equation (eq.
\ref{Ham-actionwave}) with the non-relativistic Hamiltonian, if extended to
the hybrid wavefunction $\psi(x,t)$, contains both the dynamical content for
the time evolution of the action $S(x,t)$ and all the terms in the
conservation and diffusion equation for the probability amplitude $A(x,t)$.%

\begin{equation}
i \hbar\frac{\psi(x ,t)}{ \partial t} =H\psi(x ,t) =H\left[  A(x ,t)\zeta(x
,t)\right]
\end{equation}

The connection between $i\hbar\partial\psi/\partial t$ on the left and the
dynamical quantities is given by equation (\ref{lhs-sch}). The first term
represents the single system dynamics (action-wave equation) and the matching
Hamiltonian term on the right should be the kinetic term $p^{2}/2m$,
\emph{which does not contain the quantum signature} $\hbar$. The dephasing
term in the action-wave equation is not observable, because the action-waves
are not directly observable. \emph{Dephasing manifests through the diffusion
of the probability amplitude}, as depicted in the fig. \ref{dephasing}. What
is remarkable is that all the three ensemble terms signifying the macroscopic
and microscopic stochasticity are also contained within this Hamiltonian.%

\begin{align}
\nabla^{2}\psi &  =\nabla\cdot.(A \frac{i}{\hbar} e^{i S/\hbar} \nabla S +e^{i
S/\hbar} \nabla A) =\nabla\cdot\left(  \frac{i}{\hbar} \psi S +\frac{\psi}{A}
\nabla A\right) \\
&  = -\frac{1}{\hbar^{2}} \psi\left(  \nabla S\right)  ^{2} +\frac{i}{\hbar}
\psi\nabla^{2}S +\frac{i}{\hbar} \frac{\psi}{A} \nabla A \cdot\nabla S
+\frac{\psi}{A} \nabla^{2}A +\frac{i}{\hbar} \frac{\psi}{A} \nabla A
\cdot\nabla S
\end{align}
Therefore, we get the matching terms in
\begin{equation}
-\frac{\hbar^{2}}{2 m} \nabla^{2}\psi=\frac{\psi}{2 m} \left(  \nabla
S\right)  ^{2} +\frac{ -i \hbar}{2 m} \psi\nabla^{2}S +\frac{ -i \hbar}{m}
\frac{\psi}{A} \nabla S \cdot\nabla A +\frac{ -\hbar^{2}}{2 m} \frac{\psi}{A}
\nabla^{2}A
\end{equation}

The first term is the particle momentum.\emph{\ The last three terms concern
the ensemble averaged time evolution, in which the third contains the
intrinsic uncertainty related to the quantum of action. }For a given single
event, the particle state is well defined within the small uncertainty with
the scale $\hbar$. The characteristic feature of quantum dynamics is the
Heisenberg uncertainty, encoded as the uncertainty in the phase $S/\hbar$ as
$\delta S\simeq\hbar$. This manifests in the scatter in $p$ and $x$ in the
statistical ensemble of particle events as $\Delta p\Delta x\simeq\hbar$. When
many histories of particle dynamics are averaged, the description is in terms
of the probability density, characterized by $A(x,t)$ and its derivatives.

Combining all the terms we arrive at the equation for the \textit{\ time
evolution of the ensemble averaged hybrid quantity} $\psi$,
\begin{equation}
i\hbar\frac{\partial\psi}{\partial t}=-\frac{\hbar^{2}}{2m}\nabla^{2}%
\psi+V\psi
\end{equation}
This is the Schr{\"{o}}dinger equation, which generalizes to the universal
equation for particle quantum mechanics $i\hbar\dot{\psi}=H\psi$, but with a
radically new inner structure and interpretation. In this bottom up
construction, we have discovered that the Schr\"{o}dinger equation is an
ensemble equation. It is not an equation for matter waves or probability
waves. \emph{When an ensemble of Hamiltonian action-wave dynamics is averaged,
we get the Schr\"{o}dinger equation}. Hence, quantum mechanics is derived from
the universal mechanics described by Hamiltonian action-wave mechanics.

Though Schr\"{o}dinger was inspired by Hamilton's action mechanics
\cite{Schrod}, since he developed the wave equation for de Broglie's
matter-waves, it was assumed that the equation dealt with single particles or
single dynamical histories. The severe problems of interpretation surfaced
after the equation became successful in its statistical predictions. The
reconstruction achieved by combining the ensemble averaged probability density
of particle dynamics and the action-waves shows the factual situation.

\emph{It is now clear that the particle is always in a unique physical state,
as a single whole, and not in superposition of states}. What is superposed in
single particle dynamics is the action-waves associated with the particle. The
action-waves associated with the particle states $|s_{i}\rangle$ are
\textit{present with every realization of the evolution} in the ensemble,
causing superposition and interference. $|\psi\rangle=\langle\rho\rangle
^{1/2}|s\rangle\exp(iS/\hbar)$ is the entire ensemble representation of the
physical state. This wavefunction is a mathematical entity synthesised from
the ensemble of dynamical histories and it is not a physical entity.

It may be easily verified that the ensemble averaged hybrid wavefunction with
the action-wave content is consistent with the basic axioms of quantum
mechanics. It belongs to the Hilbert space and it obeys exact Born's rule by
construction. However, being ensemble averaged, there is no `collapse of the
wavefunction' during individual measurements. The connections between the
Heisenberg formulation of quantum mechanics, Schr\"{o}dinger mechanics, the
Hamilton's equations for dynamics, and the general time evolution in terms of
Poisson brackets etc. are more transparent in terms of the new action-wave
mechanics. These relations are ensemble relations. In fact, the very situation
that these two equivalent formulations of QM were possible was due to the true
ensemble nature of the Schr\"{o}dinger equation. Some of these formal aspects
will be explored in another paper.

\subsection{Summary Comments on RQM}

\begin{figure}[ptb]%
\centering
\includegraphics[
natheight=2.415100in,
natwidth=8.875000in,
height=1.3774in,
width=5.0186in
]%
{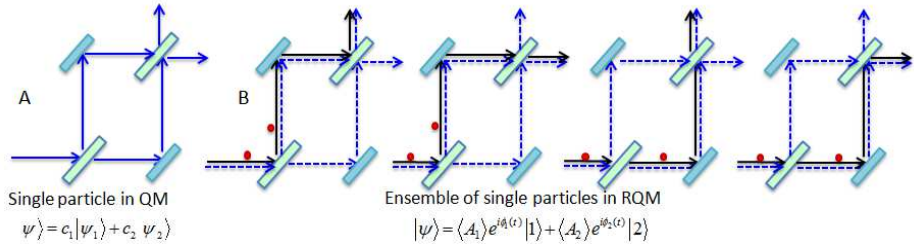}%
\caption{Matter dynamics and action-wave interference in RQM: statistical
results of measurements as well as quantum interference are correctly
reproduced, without collapse and other foundational issues. A) Conventional
QM. B) RQM: Particle dynamics has two distinct possibilities inside the
interferometer that do not superpose, and two at the exit (black arrows) The
probabilities at the exit port are determined by the relative phase of the two
interfering action-waves (broken arrows), which are present in every event. }%
\label{mach-zehnder}%
\end{figure}

Quantum mechanics is really the fusion of two fundamental features: stochastic
particle dynamics and event by event quantum interference of action-waves.
Thus, \emph{Schr{\"{o}}dinger equation is a hybrid evolution equation,
obtained as the ensemble average of the Hamiltonian action-wave equation}.
This is the crux of reconstructed quantum mechanics (RQM). All mechanics is
included in the Hamiltonian action-wave equation, and when averaged over many
dynamics histories we get the Schr{\"{o}}dinger equation, which has a ensemble
averaged `wavefunction' instead of the action-wave. The ensemble average of
stochastic particle dynamics solely determines the real positive amplitude and
the action waves are not involved. This gives exact Born's rule for the
probability density. The relative phase of all interfering action-waves
determines the probability for particle dynamics at the location of
interference. The particle is in a definite physical state at every instant
and does not partake in superposition. \emph{Both the particle and its
action-waves exist in real space and time and the hybrid wavefunction is in
the Hilbert space of QM}. I stress that the action-waves do not carry or
exchange energy or momentum. The foundational problems were already eliminated
with the Hamiltonian action-wave dynamics, and RQM is free of such birth
defects. An example of a matter interferometer in RQM and its comparison with
standard quantum mechanics are depicted in figure \ref{mach-zehnder}, as illustration.%

%TCIMACRO{\FRAME{ftbpFU}{5.0186in}{1.3774in}{0pt}{\Qcb{Matter dynamics and
%action-wave interference in RQM: statistical results of measurements as well
%as quantum interference are correctly reproduced, without collapse and other
%foundational issues. A) Conventional QM. B) RQM: Particle dynamics has two
%distinct possibilities inside the interferometer that do not superpose, and
%two at the exit (black arrows) The probabilities at the exit port are
%determined by the relative phase of the two interfering action-waves (broken
%arrows), which are present in every event. }}{\Qlb{mach-zehnder}}%
%{m-z-rqm.png}{\special{ language "Scientific Word";  type "GRAPHIC";
%width 5.0186in;  height 1.3774in;  depth 0pt;  original-width 8.875in;
%original-height 2.4151in;  cropleft "0";  croptop "1";  cropright "1";
%cropbottom "0";  filename 'M-Z-RQM.png';file-properties "XNPEU";}} }%
%BeginExpansion
%EndExpansion

We note in passing that the failure of attempts like the de Broglie-Bohm (dBB)
formulation to deal with the foundational problems can be traced to treating
the Schr\"{o}dinger equation as the fundamental equation for single particle
dynamics and the basis for interpretation, whereas factually it is the
equation for ensemble averaged dynamics. \ The unphysical features, apart from
nonlocality, of the dBB theory~can be traced to misinterpreting the ensemble
term $\left(  \hbar^{2}/2mA\right)  \nabla^{2}A$ as a quantum potential acting
on \emph{single particle dynamics}. In fact, such attempts took us farther
from the right path because the dBB approach was trying to restore determinism
to quantum mechanics, with position as hidden variable and the momentum
through the action equation $\partial S/\partial x=p$, both specified, whereas
the complete solution was in the action-wave dynamics with universal
indeterminism. More details may be found in appendix 3.

\subsection{Indeterminism as a Fundamental Feature}

We saw in the road to the reconstruction that there is inherent indeterminism
in mechanics because particle dynamics is governed by the action-wave and the
fundamental (minimum) scale of action in the action-wave, $\varepsilon
=\hbar\text{.}$ This results in the dephasing of the action-waves associated
with particle mechanics. Such indeterminism is not specific to quantum
mechanics; it is part of universal mechanics, including classical
(macroscopic) mechanics, independent of the physical scales and the form of
Hamiltonian. Dephasing of action-waves manifests as the (dissipation-less)
diffusion of the probability density. All dynamics is indeterministic, at the
scale of action $\hbar\text{.}$ That is the fundamental lesson from the
reconstruction. When $\varepsilon\nabla^{2}S\ll\left(  \nabla S\right)  ^{2}$,
dynamics according to the approximate Hamilton-Jacobi equation cannot be
distinguished from the dynamics of the complete Hamiltonian action-wave
equation. Then the small statistical spread in the different dynamical
histories cannot be discerned. Similar condition is used for the WKB
approximation in quantum mechanics,
\begin{equation}
\left(  \frac{\hbar}{p^{2}}\frac{\partial p}{\partial x}\right)  \ll1
\end{equation}
Here, we see clearly that the celebrated Hamilton-Jacobi equation and the
dynamics it represents (including Newton's equation of motion) are
approximations to a more complete classical mechanics described by the
action-wave equation. The determinism in classical mechanics is only apparent,
and not fundamental.

Since the wavefunction is the result of ensemble average of the histories
particle dynamics and the action-waves, I have cleared the mystery why the
Schr\"{o}dinger equation is a differential equation for the probabilities of
observables, while accurate on the interference and correlation effects. The
long held hope and search for a deterministic dynamics turn out to be a
mirage-chase. \emph{Indeterminism is not a foundational problem; it is a
foundational feature}.

\subsection{Quantum Information}

The reconstructed mechanics makes it clear that a single quantum system, like
a two-level atom, in factually in only one of the states at any instant, and
not in a superposition of the two states. What are superposable are the
action-waves corresponding to the states. Thus, the physical aspect of a
quantum-bit is very different in RQM, compared to conventional QM. All
interference effects that enable quantum information processing and quantum
computing are traceable to the action-waves. As stated before, entanglement of
particle states is an ensemble statement. The superposition represented as
$c_{1}|+\rangle_{1}|-\rangle_{2}+c_{2}|-\rangle_{1}|+\rangle_{2}$ is
applicable only to the associated action-waves. Quantum information processing
then deals with the physical states of matter and their evolution, as
described by the action-wave dynamics; it does not have any special role or
physical priority in the understanding of matter and its dynamics.

\subsection{Fundamental Quantum Scales}

The reconstruction of mechanics that I outlined has one unique scale or
quantum of action, which is empirically determined as $\hbar\text{.}$ There is
no additional quantization scales associated with the dynamical quantities
like energy and momentum, or with the coordinate variables like position. Even
the fundamental quantization of spin is in fact just the quantization of
action arising in the `looped' nature of the action-waves. This is significant
when we consider speculations based on synthesized scales like the Planck
quantities. In RQM, such scales are not expected to be fundamental. RQM
predicts that there is no quantum to classical transition associated with the
scale of Planck mass or any macroscopic mass scales. Since there is no quantum
measurement problem and the notion of the collapse of the wavefunction in RQM,
speculations of new physics in quantum mechanics at some such scale associated
with mass or gravity are rendered redundant. In particular, the speculations
on spontaneous localization of the wavefunction are no more relevant, because
the wavefunction is a hybrid mathematical entity and not a physical entity in
space and time.

The other Planck scales that are frequently mentioned are the Planck length
and Planck time, both being in the extreme microscopic compared to what is
practically accessible. RQM based on the single scale of action does not
expect new quantum effects at these scales.

The de Broglie relations for the hypothetical matter waves were based on
applying the optical wave relations to the relativistic energy of a particle.
Then, $E=\hbar\omega$ and $p=\hbar k=h/\lambda$. The phase velocity of de
Broglie's matter waves is $v_{p}=E/p=c^{2}/v$, where $v$ is the velocity of
the particle. The group velocity $d\omega/dk$ coincides with the particle
velocity. The action-wave relations in the modified Hamiltonian action
dynamics are different, and physically more reasonable. The action relations
are
\begin{equation}
\frac{\partial S}{\partial t}=-H,\text{\quad}\frac{\partial S}{\partial x}=p
\end{equation}
Both have the additional terms arising due to the fundamental scale of action
$\hbar$. This defines a time and length scales for the action-waves $\delta
t_{0}=h/H$ and $\delta x_{0}=h/p$. The action has multiple contributions
including those unrelated to motion. These terms must be subtracted to get the
terms related to true motion, to find the ratio $E/p$. The action contribution
from the rest mass-energy is one such. Thus, both the phase velocity and group
velocity are below $c$ in RHAM.

\subsection{Divergence of Zero-Point Energy}

A serious problem with quantum mechanics arises when we consider
multi-particle wavefunctions and its limit in a quantum field theory, like
quantum electrodynamics or even quantum optics. Since the conventional `waves'
in quantum mechanics (associated with real photons) have energy and momentum,
even the wave modes of the electromagnetic vacuum have the zero-point energy
of $\hbar\omega/2$ per wave mode of each polarisation. This results in
divergent energy density in space, in clear contradiction with the observed
slow dynamics of the universe, accurately determined in the Hubble parameter.
This problem of the very large `cosmological constant' and such divergences in
quantum fields are solved in the modified Hamiltonian action dynamics because
the waves in reality are the action-waves that do not have energy or momentum.
The particle states represent a finite number of particles with a finite
zero-point energy, which is in fact negligible. This is a significant
conceptual advance with further implications to quantum field theories and the
interpretation of second quantization. In particular, there are no vacuum
modes with zero-point energy in quantum optics based on RQM. This is
well-supported by recent homodyne experiments done in our laboratory in a
configuration that avoided the conventional symmetric beam splitter in the
balanced homodyne set up \cite{Homodyne}.

\subsection{General Relativity and Action-Wave Mechanics}

The perceived incompatibility between the general theory of relativity (GTR)
and quantum mechanics is well known \cite{Ashtekar}. This is a serious
impediment to completing the quantum theory of gravitational phenomena. The
new action-wave mechanics changes this grim scenario positively. First of all,
Schr\"{o}dinger equation is not the equation for each dynamical history; it is
an ensemble equation. Therefore, the incompatibility of the two theories is
not determined by the Schr\"{o}dinger equation. Since I have shown that the
Hamiltonian action mechanics has the universal irreducible uncertainty,
general relativity that obeys the principle of stationary action will not be
an exception. GTR will necessarily inherit this microscopic uncertainty in
action, related to the fundamental scale $\hbar$, in a suitable modification.
The variational principle is applicable to gravitational physics through the
geometric action that contains the Ricci scalar curvature $R$ and the
determinant of metric $g_{\alpha\beta}$,%
\begin{equation}
S_{GR}=\frac{c^{4}}{16\pi G}\int R\sqrt{-g}d^{4}x
\end{equation}
With the modification to Hamiltonian mechanics, the time evolution of this
action is modified as well, with the additional microscopic term involving
$\hbar$. The major conceptual change is that matter-energy and its
gravitational fields corresponding to different physical states do not
superpose. This is applicable irrespective of whether the particle is treated
as an extended entity or not. The action-waves that superpose do not carry
potentially divergent energy-momentum or its gravity. With the seemingly
impenetrable wall between the `classical' and the quantum world dissolved, and
the quantum features of superposition confined to the action-waves, a quantum
theory of gravity looks more feasible than ever \cite{Unni-QG}.

But this needs conceptual revisions at two different fronts. One is to
reconstruct general relativity to incorporate the action-wave basis, as for
any classical dynamical theory that respects the principle of stationary
action. This will integrate the irreducible uncertainty of the quantum of
action with general relativity, already providing the universal equation for
single dynamical histories. Due to the tiny magnitude of the quantum of
action, none of the empirical results of GTR would be affected. We have seen
that the matter states and the states of gravitational fields do not form
superpositions; only related action-waves can be in superposition. Hence,
there is no more the speculated superposition of `geometries'. Also, the
Schr\"{o}dinger equation deals with the ensemble dynamics, and not single
quantum histories. This new feature needs to be formally incorporated.

The second front involves moving closer to true physical dynamical variables
rather than working in an effective geometric description. In
electromagnetism, the effect of the sources (charges and their currents) is
described in terms of their potential fields. For gravity, it is the effect of
such potential fields on spatial and temporal intervals that are taken as the
primary dynamical variable. (Electromagnetism does not affect neutral matter,
whereas gravity affects all matter with mass-energy, with `equivalence'). As a
consequence, the gravitational potentials and the space-time intervals between
matter states are identified. However, the metric-references are invariably
constructed out of matter. There is no clock or physical time without matter.
Also, there is only one matter-energy filled universe where all physical
phenomena are actualized. It is easy to prove, with both logical arguments and
several experimental results, that such a universe has the role of a master
determining frame \cite{Unni-ISI}. Then GTR itself and Einstein's equations
need a crucial modification to make the theory consistent with its singular
relevance in this universe \cite{Unni-Moriond}. This `Centenary Einstein's
Equations' (CEE) is
\begin{equation}
R_{ik}-\frac{1}{2}g_{ik}R-\frac{8\pi G}{c^{4}}U_{ik}=\frac{8\pi G}{c^{4}%
}T_{ik}%
\end{equation}
The energy-momentum tensor of the matter-energy in the universe, $U_{ik}$, is
the integral part of this equation. Therefore, the possibility of the vacuum
Einstein's equations has disappeared. Apart from being in complete agreement
with all experimental tests and observations, this equation sacrifices general
coordinate invariance to gain many new predictions \cite{Unni-ISI,Unni-CIRI},
consistent with the fact that all observable gravitational phenomena unfolds
in this one, and only one, universe.

I believe that the road to a theory of quantum gravity is straighter and
clearer with Hamiltonian action-wave mechanics, RQM, and CEE as the basis.

\section*{Concluding remarks}

I have reconstructed and generalized Hamiltonian action mechanics
incorporating the action-waves that are responsible for the universal validity
of the principle of action in a new action-wave equation. This is the
universal equation for single dynamical histories and it contains the
microscopic irreducible uncertainty arising from a fundamental quantum of
action. This key step solves all vexing foundational and conceptual problems
of quantum mechanics. The theory of quantum mechanics based on the
Schr\"{o}dinger equation turns out to be an emergent theory that results from
ensemble averaging the action-wave dynamics. The wavefunction of the
Schr\"{o}dinger equation is a hybrid entity encoding the ensemble averaged
stochastic dynamics of particles with distinct physical states and the
action-waves that accumulate the dynamical phases in all possibilities, event
by event. There are no matter-waves. The new wave-particle connection, rather
than duality, in the action-wave mechanics eliminates the collapse of the
wavefunction, the measurement problem, and the quantum-classical divide. It
restores separable independent particle states in quantum entanglement and
eliminates nonlocality. The reconstruction extends to relativistic QM, quantum
optics, and the Dirac equation, enabling the resolution of some additional
foundational issues. With the action-waves as phase carriers, the divergent
energy that is usually associated with the quantum vacuum modes is no more
troublesome. It is hoped that RHAM and the emergent RQM will result in the
definite closure of the philosophical debates anchored on the conceptual
issues of QM~\cite{StanfordQM}. RHAM makes it clear that matter-energy states
and their long-range fields do not superpose. The entity in superposition, the
action-waves, do not have energy-momentum, nor associated fields. In
particular, the `classical' theory of gravity, which respects the action
principle, will need the same Hamiltonian action-wave modification and the
ensemble average of such a theory will be important for the theory of quantum
gravity. We note that there is no physical sense in the ensemble average over
dynamical histories when there is only a single realisation of a material
system, like the universe. Therefore, RHAM will rechart the conceptual
trajectory in the fields of quantum gravity and quantum cosmology.

\bigskip

\section*{Acknowledgments}

I thank Daksh Lohiya and S. Sankaranarayanan for several discussions on the
path towards completing this work, and Martine Armand for help in clear
detailing of this paper.

\bigskip\appendix

\section*{Appendix 1: Massive Neutrinos}

The discovery that neutrino flavours show quantum interference effects, by
which the probability to detect them in any particular flavour `oscillates',
is explained in conventional quantum mechanics by representing each flavour
state $\left(  \left\vert \nu_{e}\right\rangle ,\left\vert \nu_{\mu
}\right\rangle ,\left\vert \nu_{\tau}\right\rangle \right)  $ \ as a linear
superposition of three mass eigenstates $\left(  m_{1},m_{2},m_{3}\right)  $.
Thus, the physical picture of a neutrino that is emitted in a nuclear reaction
is as a particle propagating as a superposition of three mass states. This
implies that each neutrino propagates with a superposition of 3 distinct
gravitational fields. Modified action-wave mechanics and RQM clarifies that
the quantum mechanics of individual neutrinos is factually very different from
this picture. Every neutrino, irrespective of its flavour, is in one of the
mass eigenstates and not in a superposition. This (random) state is retained
from emission to detection. This also implies that the gravitational field of
a particular neutrino corresponds to only one of the possible masses, and not
a superposition of three possibilities of fields or geometries. The
action-waves corresponding to the three mass-energy states copropagate with
the neutrino and generate all interference effects. When observed, the
probability to be in each flavour is determined by the local relative phase of
the action-waves, exactly similar to the case of an interferometer. I
illustrate the `neutrino oscillation' phenomena in RQM, with the simpler case
of a two neutrino flavours $\left(  \left\vert \nu_{e}\right\rangle
,\left\vert \nu_{\mu}\right\rangle \right)  $ and two mass states $\left(
m_{1},m_{2}\right)  $.

Consider a neutrino generated as an electron neutrino $\nu_{e}$, which has two
possible mass states $m_{1}$ and $m_{2}$. The neutrino particle is randomly in
one and only one of these mass states, say $m_{2}$ and not in a superposition
of states. The action-waves associated with $m_{1}$ and $m_{2}$ are
$\left\vert 1\right\rangle \exp(iE_{1}t/\hbar)$ and $\left\vert 2\right\rangle
\exp(iE_{2}t/\hbar)$ in the proportion determined by the details of particle
physics (mixing angles). Their interference causes periodic changes in the
relative phase as $\exp i\left(  E_{1}-E_{2}\right)  t/\hbar$. The action-wave
superposition at time $t$ is
\begin{equation}
\left\vert \nu(t)\right\rangle =\left\vert 1\right\rangle c_{m1}e^{i\omega
_{1}t}+\left\vert 2\right\rangle c_{m2}e^{i\omega_{2}t}%
\end{equation}
$\left\vert c_{m1}\right\vert ^{2}+\left\vert c_{m2}\right\vert ^{2}=1$ and
$\omega_{i}=E_{i}/\hbar$. The probability to remain an electron neutrino when
observed at time $t$ is
\begin{align}
P_{ee}  &  =\left\vert \left\langle \nu_{e}\right\vert \left\vert
\nu(t)\right\rangle \right\vert ^{2}=\left\vert \left\vert c_{m1}\right\vert
^{2}e^{i\omega_{1}t}+\left\vert c_{m2}\right\vert ^{2}e^{i\omega_{2}%
t}\right\vert ^{2}\\
&  =\left\vert c_{m1}\right\vert ^{4}+\left\vert c_{m2}\right\vert
^{4}+\left\vert c_{m1}\right\vert ^{2}\left\vert c_{m2}\right\vert ^{2}\left(
e^{i\left(  \omega_{2}-\omega_{1}\right)  t}+e^{-i\left(  \omega_{2}%
-\omega_{1}\right)  t}\right) \nonumber\\
&  =1-2\left\vert c_{m1}\right\vert ^{2}\left\vert c_{m2}\right\vert
^{2}+2\left\vert c_{m1}\right\vert ^{2}\left\vert c_{m2}\right\vert ^{2}%
\cos\left(  \omega_{2}-\omega_{1}\right)  t\\
&  =1-2\left\vert c_{m1}\right\vert ^{2}\left\vert c_{m2}\right\vert
^{2}\left[  1-\cos\left(  \omega_{2}-\omega_{1}\right)  \right]  t\nonumber\\
&  =1-4\left\vert c_{m1}\right\vert ^{2}\left\vert c_{m2}\right\vert ^{2}%
\sin^{2}\left(  \omega_{2}-\omega_{1}\right)  t/2
\end{align}

The probability of the electron neutrino to be detected as a muon neutrino is
\begin{equation}
P_{e\mu}=\left\vert \left\langle \nu_{\mu}\right\vert \left\vert
\nu(t)\right\rangle \right\vert ^{2}=1-\left\vert \left\langle \nu
_{e}\right\vert \left\vert \nu(t)\right\rangle \right\vert ^{2}=4\left\vert
c_{m1}\right\vert ^{2}\left\vert c_{m2}\right\vert ^{2}\sin^{2}\left(
\omega_{2}-\omega_{1}\right)  t/2
\end{equation}
Since
\begin{equation}
E=\left(  p^{2}c^{2}+m^{2}c^{4}\right)  ^{1/2}=pc\left(  1+\frac{m^{2}c^{4}%
}{2p^{2}c^{2}}\right)  \simeq E+\frac{m^{2}c^{4}}{2E}%
\end{equation}
We have $\left(  \omega_{2}-\omega_{1}\right)  t=\frac{c^{4}}{2\hbar E}\left(
m_{2}^{2}-m_{2}^{2}\right)  \frac{L}{c}=\frac{c^{3}}{2\hbar E}\Delta m^{2}L$.
If $c_{m1}=\cos\theta$ and $c_{m2}=\sin\theta$, as conventionally coded, then
\begin{equation}
\left\vert \left\langle \nu_{\mu}\right\vert \left\vert \nu(t)\right\rangle
\right\vert ^{2}=4\sin^{2}\theta\cos^{2}\theta\sin^{2}\left(  \frac{c^{3}%
}{4\hbar E}\Delta m^{2}L\right)  =\sin^{2}2\theta\sin^{2}\left(  \frac{c^{3}%
}{4\hbar E}\Delta m^{2}L\right)
\end{equation}

We see that the particular neutrino particle that started out as \ an electron
neutrino remained in a single mass state (with mass $m_{2}$ here) throughout
its propagation. The interference of action-waves corresponding to the two
mass states determined the probability of detecting the neutrino in either of
the flavour states at time $t$ after emission. Another electron neutrino might
be in mass state $m_{1}$ or $m_{2}$, randomly and with probabilities $p_{1}$
and $p_{2}$. In RQM, $\left\vert c_{m1}\right\vert =\sqrt{p_{1}}$ and
$\left\vert c_{m2}\right\vert =\sqrt{p_{2}}$.

\section*{Appendix 2: The Dirac Equation}

Quantum mechanics reconstructed from the universal action-wave mechanics
naturally extends to more general Hamiltonians. The dynamical equation of the
particle is always given by the time evolution of the action wave
$\zeta(x,t)=\exp(iS/\hbar)$,
\begin{equation}
i\hbar\frac{\partial\zeta}{\partial t}=H\zeta
\end{equation}
Therefore, the extension to the ensemble averaged wavefunction is a
\emph{first order time evolution equation, for any Hamiltonian}.

Consider a non-relativistic example of a spin-1/2 particle with a magnetic
moment. The magnetic energy in an applied field $B$ is $E_{m}=\mu_{0}s\cdot
B\text{,}$ where the spin $s$ is two-valued. We have the two-valued energy
\begin{equation}
E_{m}=\pm\mu\left\vert B\right\vert =\pm\frac{\mu_{0}}{2}\left(  B_{x}%
^{2}+B_{y}^{2}+B_{z}^{2}\right)  ^{1/2}%
\end{equation}
As well known in the context of the Pauli equation (generalized from the
Schr\"{o}dinger equation), $i\hbar\frac{\partial\psi_{i}}{\partial t}%
=H_{ij}\psi_{j}$, the linear Hamiltonian corresponding to this is
\begin{equation}
H=\mu_{0}\sigma\cdot B=\mu_{0}\left(  \sigma_{x}B_{x}+\sigma_{y}B_{y}%
+\sigma_{z}B_{z}\right)
\end{equation}
where $\sigma_{i}$ are the Pauli $2\times2$ matrices. Explicitly written,
\begin{equation}
H_{ij}\psi_{j}=\mu_{0}\left[
\begin{array}
[c]{cc}%
B_{z} & B_{x}-iB_{y}\\
B_{x}+iB_{y} & -B_{z}%
\end{array}
\right]  \left[
\begin{array}
[c]{c}%
\psi_{u}\\
\psi_{d}%
\end{array}
\right]
\end{equation}

We are also familiar with the common physical situation in which a particle in
a symmetric double well potential has the tunneling probability characterized
by a frequency $\Omega$ and the energy scale $\hbar\Omega\text{.}$ The bare
energy levels split into a doublet $\pm\hbar\Omega\text{.}$ Then we have
\begin{equation}
H_{mn}\psi_{n}=\hbar\left[
\begin{array}
[c]{cc}%
\Omega & 0\\
0 & -\Omega
\end{array}
\right]  \left[
\begin{array}
[c]{c}%
\psi_{+}\\
\psi_{-}%
\end{array}
\right]
\end{equation}

Now I consider the spin-1/2 particle in a magnetic field, in the double well
potential with tunneling probability characterized by $\hbar\Omega\text{.}$
The energy levels split such that the total energy is
\begin{equation}
E_{\pm}=\pm\left(  \hbar^{2}\Omega^{2}+\mu^{2}B_{x}^{2}+\mu^{2}B_{y}^{2}%
+\mu^{2}B_{z}^{2}\right)  ^{1/2} \label{split-energy}%
\end{equation}
Because of the independent two-valuedness of the two physical situations, and
the linearity of action-wave mechanics, the Hamiltonian now have to be
4-dimensional, with a 4-component wavefunction. We can write it by inspection,
to get the correct eigenvalues of the energy $E_{\pm}\text{.}$ We get
\begin{equation}
H_{ij}\psi_{j}=\left[
\begin{array}
[c]{cccc}%
\hbar\Omega & 0 & \mu_{0}B_{z} & \mu_{0}(B_{x}-iB)\\
0 & \hbar\Omega & \mu_{0}(B_{x}+iB) & -\mu_{0}B_{z}\\
\mu_{0}B_{z} & \mu_{0}(B_{x}-iB) & -\hbar\Omega & 0\\
\mu_{0}(B_{x}+iB) & -\mu_{0}B_{z} & 0 & -\hbar\Omega
\end{array}
\right]  \left[
\begin{array}
[c]{c}%
\psi_{1}\\
\psi_{2}\\
\psi_{3}\\
\psi_{4}%
\end{array}
\right]  \label{Dirac-form}%
\end{equation}
We cannot get the right energy expression with a $2\times2$ Hamiltonian
because we need to maintain the two-valuedness of energy for any linear
combination of the spin projections. We can arrive at this Hamiltonian more
formally by demanding that the total energy should match the expression for
the linear Hamiltonian. Extending the Pauli Hamiltonian to include the
tunneling, we get
\begin{equation}
H=\mu\alpha\cdot B+\beta\hbar\Omega
\end{equation}
where the matrices $\alpha$ are similar to the Pauli matrices and the matrix
$\beta$ incorporates the tunneling term. The total energy should match the
expression for the Hamiltonian,
\begin{equation}
\hbar^{2}\Omega^{2}+\mu^{2}\left(  B_{x}^{2}+B_{y}^{2}+B_{z}^{2}\right)
=\left(  \mu\alpha\cdot B+\beta\hbar\Omega\right)  \cdot\left(  \mu\alpha\cdot
B+\beta\hbar\Omega\right)
\end{equation}
This determines the the matrices $\alpha$ and $\beta\text{.}$ We get
\begin{align}
\alpha_{i}\alpha_{j}+\alpha_{j}\alpha_{i}  &  =2\delta_{ij}\\
\beta^{2}  &  =1;\quad\alpha_{i}\beta+\beta\alpha_{i}=0
\end{align}
The anti-commuting nature originates in the requirement that the energy has
two independent contributions, without cross terms. \emph{This
non-relativistic Hamiltonian has the same structure as the Dirac Hamiltonian
and the matrices }$\alpha_{i}$\emph{ and }$\beta$\emph{ are the Dirac
matrices}. We see at once that it is Hamiltonian action-wave dynamics and not
relativity that determines the structure and form of the Dirac equation, even
though, historically, relativity was the motivating force. The equation for
dynamics of the particle remains identical in all cases,
\begin{equation}
i\hbar\frac{\partial\zeta}{\partial t}=H\zeta\rightarrow i\hbar\frac
{\partial\psi}{\partial t}=H\psi
\end{equation}
Only the form of the Hamiltonian changes.

The particle in RQM tunnels back and forth, stochastically, without being in a
superposition. The probability is determined by the relative phase of the
action-waves, as usual. When $B=0\text{,}$ the action-waves corresponding to
the two energy states are $\exp(i\Omega t)$ and $\exp(-i\Omega t)\text{.}$ The
ensemble averaged symmetric wavefunction with its amplitude as $A=\sqrt{\rho}$
is
\begin{equation}
\psi(x,t)=\frac{1}{\sqrt{2}}\left(  \chi_{+}\exp\left(  i\Omega t\right)
+\chi_{-}\exp\left(  -i\Omega t\right)  \right)
\end{equation}
As I demonstrated for other examples, the particle is not in a superposition
of two states; only the action-waves are.

The expression for the total energy in equation (\ref{split-energy}) reduces
to
\begin{equation}
E_{\pm}=\pm\hbar\Omega\left(  1+\frac{\mu^{2}B^{2}}{2\hbar\Omega}\right)
\end{equation}
in the limit of small magnetic field, $\left\vert \mu B\right\vert \ll
\hbar\Omega\text{.}$ Similarity to the expression for non-relativistic kinetic
energy $p^{2}/2m$ is evident. One may check that the solutions split into a
large component $\psi_{L}$ and a small component $\psi_{S}\sim\frac{\mu
^{2}B^{2}}{2\hbar\Omega}\psi_{L}\text{,}$ as it happens in the case of the
Dirac equation.

The Dirac equation for the relativistic electron is
\begin{equation}
i\hbar\frac{\partial\psi}{\partial t}=H\psi
\end{equation}
where the Hamiltonian corresponds to the relativistic energy
\begin{equation}
E=\pm\left(  m^{2}c^{4}+p^{2}c^{2}\right)  ^{1/2}%
\end{equation}
The relevant Hamiltonian is similar to the non-relativistic Hamiltonian we
considered,
\begin{equation}
H=\alpha\cdot pc+\beta mc^{2}%
\end{equation}

Since the particle is never in a superposition of different states, the
factual situation in RQM in relativistic mechanics does not involve
difficulties of conventional Dirac quantum mechanics, like `zitterbewegung'.
The interferences are between the action-waves and not between the
(non-existent) matter-waves. The relative phases are relevant for the
probabilities during observation and not for the actual dynamics of the
particle. (It is significant that the zitterbewegung motion was highlighted by
Schr\"{o}dinger, who was a strong advocate for real matter-waves). In summary,
the `zitter motion' is an ensemble result and a single particle phenomenon.
Analogue experiments confirm this as an ensemble averaged observable quantity
\cite{zitter-ion}.

\section*{Appendix 3: A Note on Bohmian Mechanics}

The differences between the robust basis of action-wave mechanics and all
previous paths taken to understand quantum mechanics, especially
interpretational ones, are transparent. All attempts to resolve the
foundational issues and to get a consistent and rational interpretation
mistook the Schr\"{o}dinger equation and the wavefunction as describing the
single dynamical history. Einstein's departure from this universal view did
not go far either. RQM actualises what was crucially missing in Einstein's
speculation on the statistical interpretation of the wavefunction. A purely
statistical interpretation of the wavefunction cannot reproduce single
particle interference. In RQM, the real amplitude part of the wavefunction is
statistical and ensemble averaged, while the action-wave can remains invariant
(coherent over the ensemble). Thus, the empirically and conceptually crucial
single particle interference is reproduced in RQM. Since matter itself \ has
no wave property in RQM, it is obviously different from all approaches that
identify the Schr\"{o}dinger wavefunction with any single history with
ontological status in space and time or in configuration space.

Since the basis of action-wave dynamics refers to the universal method
invented by Hamilton, expressed through the action-wave $\zeta(x,t)=\exp
(iS(x,t)/\hbar)$, there is the possibility that RQM may be confused with a
totally different approach named after L. de Broglie and D. Bohm that uses the
Schr\"{o}dinger wave function in the polar form $\psi_{dBB}=R\exp(iS/\hbar)$
\cite{Bohm-review,Holland}. It is obvious that the functions $\zeta(x,t)$ and
$\psi_{dBB}$ are very different. The de Broglie-Bohm (dBB) theory is based on
analysing the Schr\"{o}dinger equation as the fundamental equation for single
particle dynamics. I have already shown that this is factually incorrect and
that the Hamiltonian action-wave equation is the fundamental equation.
Schr\"{o}dinger equation and its wavefunction pertain to the ensemble. Since
the dBB theory uses this ensemble wavefunction to calculate what it calls the
`quantum potential' that controls the single particle dynamics in the theory,
there is an inadvertent conceptual mix-up that renders the theory physically deficient.

The dBB theory splits the Schr\"{o}dinger equation into its constituents of
action dynamics and continuity of probability density. The dynamical equation
is the classical Hamilton-Jacobi equation $\partial S/\partial t=-H$, now with
a nonlocal quantum potential $V_{q}=-\left(  \hbar^{2}/2mR\right)  \nabla
^{2}R$. This is the vestige of wrongly taking the Schr\"{o}dinger equation as
the single particle equation; thus the ensemble term is misinterpreted as a
potential (without sources) for single particle dynamics! Central to the dBB
approach is the real existence of the wavefunction in the multi-particle
configuration space for each individual history of the quantum evolution.
Then, of course, the issues of nonlocality, inseparability etc. are inherent
the dBB theory. This alone shows the drastic difference between RQM and dBB approach.

The dBB theory uses position of the particle as hidden variables and a
classical guiding equation that specifies the velocity of each particle. Thus,
with both position and velocity specified, there are real trajectories in dBB
scheme, as in classical mechanics. Positions are given as initial conditions
and the velocity in terms of $\nabla S$. However, these trajectories are
controlled by the quantum potential $V_{q}=-\left(  \hbar^{2}/2mR\right)
\nabla^{2}R$ defined by the whole nonlocal wavefunction $R\exp(iS/\hbar)$. One
can either work with the quantum potential or directly in terms of the
equation $v_{j}=-i\hbar/2m_{j}\rho\left(  \psi^{\ast}\nabla_{j}\psi-\psi
\nabla_{j}\psi^{\ast}\right)  $, but in either case the probability density
given by the whole wavefunction nonlocally determines each single particle
velocity. \emph{The identifying feature of Bohmian approach is nonlocality},
as Bohm stressed \cite{Bohm-review}. RQM has no such irrational features
because the wavefunction, being an ensemble averaged abstract hybrid
construct, has no role in the single particle behaviour. RQM has no guiding
equation and no classical trajectories. In fact, RQM emerged from the basic
assertion that even classical mechanics has inherent indeterminism specified
by the Hamiltonian action-wave equation. RQM respects the uncertainty relation
at the single particle level, arising in the quantum of action; thus the
position and momentum of the particle cannot be specified simultaneously.

The failure of the dBB approach, traced to misinterpreting an ensemble
quantity as a single system potential $V_{q}$, can be illustrated vividly with
an example that was \ originally pointed out by Einstein in 1953
\cite{Einstein-BornFesch}. A particle in a definite energy state in a 1-d box
$\left(  -l,+l\right)  $ has the Schr\"{o}dinger wavefunction as a
superposition of two harmonic waves moving in opposite directions,
\begin{equation}
\psi(x,t)=\frac{1}{2}Ae^{-i\omega t}\left(  e^{iax}+e^{-iax}\right)
=Ae^{-i\omega t}\cos ax
\end{equation}
The box is assumed to be much larger than the de Broglie wavelength. All the
statistical results of standard quantum mechanics are consistent, in the
statistical sense. However, the dBB approach has a glaring inconsistency even
in this simple instance, because of its mixing the Schr\"{o}dinger
wavefunction and the single particle dynamics. The central guiding equation
gives $p/m=\partial S/\partial x=a-a=0$. At all positions, \emph{the particle
does not move} - neither to the left nor to the right! This is independent of
whether the particle is microscopic or macroscopic. Then, when observed, the
wavefunction and the quantum potential (which holds the particle stationary)
change instantaneously, allowing the particle to move at $p/m=\pm a$. The
acceleration involved can be arbitrarily large. Thus, though the problem of
collapse and conflict with Born's rule is avoided in dBB, the particle
behaviour is still discontinuous and more bizarre than in standard QM
(standard QM bypasses the issue by dealing with only statistical averages in observations).

The situation with a pair of \textquotedblleft entangled
particles\textquotedblright\ shows the difference even more clearly. The dBB
wavefunction, while dealing with a particular pair, is the same entangled
wavefunction of standard QM with its nonlocal nature persistent over arbitrary
spatial separation. Now the wavefunction depends on the coordinates of both
the particles. A measurement on one particle affects the dynamics of the other
through the nonlocal quantum potential. There is no such irrational feature in
RQM, as we have already shown.

\section*{Appendix 4: The EPR Argument}

The Einstein-Podolsky-Rosen argument on the incompleteness of the wavefunction
description in QM has an `EPR version', as published in the article in
Physical Review \cite{EPR}, and a more clear and concise `Einstein version',
available in print through his many articles \cite{Dialectica} and also
published letters (to E. Schr\"{o}dinger, M. Born, K. Popper etc.). I present
the essential argument about the lack of one-to-one correspondence between the
quantum mechanical state represented by the wavefunction and the physical
state of material particles. Consider two spin-1/2 particles that have moved
apart into space-like separated regions after splitting from the spin-zero
composite state.

Let us assume that the particles are in some unknown \emph{physical state} of
spin. \emph{We do not assume that spin projection values in more than one
direction can be specified}. This is very important for avoiding the common
confusion about the correct EPR argument (see below). In other words, we do
not assume that the particle has definite spin projections in multiple
directions simultaneously, as in classical theories and classical hidden
variable theories. So, we do not make any assumption that violates
non-commutativity of spin projections along different directions. To stress
again, we just assume that particles have some physical state in which some
observables possibly may have definite values, without violating the quantum
mechanical restrictions on non-commuting observables. Hence, there is no
assumption or mention of `physical reality' or `elements of reality'. (These
notions are superfluous in the core argument as Einstein himself clarified
\cite{Dialectica}. According to him the \textquotedblleft main argument was
buried in erudition\textquotedblright).

The joint quantum mechanical state of the particles is specified by a
wavefunction. In the basis $|x\rangle\text{,}$ it is given by
\begin{equation}
\psi_{1,2}=\frac{1}{\sqrt{2}}|+x,-x\rangle-|-x,+x\rangle
\end{equation}
The same state can also be represented in the non-commuting basis $|y\rangle$
of the orthogonal y-direction or in any basis $n$ as
\begin{equation}
\psi_{1,2}=\frac{1}{\sqrt{2}}|+n,-n\rangle-|-n,+n\rangle
\end{equation}
A measurement of the spin projection on either particle will give random
results $\pm\text{,}$ but the results on the two particles will be in perfect
anti-correlation, pair by pair, if measured along the same direction.

Now I\emph{\ state the only crucial assumption in the analysis, of Einstein
locality}. The physical state of one particle cannot be influenced or changed
by a measurement on the other spatially separated particle. I stress that the
assumption is about the physical state and not about its representation in
quantum mechanics by a wavefunction. The QM state can of course be changed by
a distant measurement. If the spin is measured along the any arbitrarily
chosen direction on particle 1, the result will be either $+$ or $-\text{.}$
The wavefunction of the particle after the measurement is definitely
$|+n_{1}\rangle$ (or $|-n_{1}\rangle$ ). The corresponding wavefunction of the
other particle on which no measurement has been done becomes $|-n_{2}\rangle$
(or $|+n_{2}\rangle$ ). Since the direction $n$ is arbitrary, two different
choices for $n$ correspond to two different wavefunctions, and distinct QM
physical states. This implies that the quantum mechanical state of a particle
can be changed by a measurement on another particle in a spatially separated
region. However, we have assumed that the physical state cannot be affected in
a spatially separated region. Now there is a conflict; the physical state
cannot change, but the quantum mechanical state can be forced to be one of the
allowed states in a basis of free choice. Therefore, there is no one-to-one
correspondence between the physical state and the QM state. Hence, QM is incomplete.

That is the precise content of the Einstein version of the EPR argument.
Nowhere in this proof, Einstein assumed that the physical state of the
particle is specified by definite values for non-commuting observables. On the
contrary, I explicitly stated that the physical state respects the
restrictions of non-commutativity.

If there is any lingering doubt about the strength of this proof, I can state
a \emph{new version that is stronger}. Going back to the form of the entangled
state
\[
\psi_{1,2}=\frac{1}{\sqrt{2}}|+x,-x\rangle-|-x,+x\rangle
\]
we note that \emph{neither particle has its own physical state whatsoever,
within QM description}. For, the general state is $a|x+\rangle+b|x-\rangle$
for the first particle and $c|x+\rangle+d|x-\rangle$ for the second. Since the
entangled state is not the product of these general states, and since there no
single particle states other than these, neither particle has a quantum state.
Measurement on one particle endows it with a state, and then \emph{the
spatially separated unobserved particle gets a definite QM state from a
situation it has no state whatsoever}. Therefore, QM states can be changed
factually by measurements in a spatially separated region. So, we have proved
that wavefunction description of the physical state is not compatible with
Einstein locality. If, on the other hand, we insist that the QM state is the
faithful representation of the physical state, Einstein locality is violated.
This proof does not need any additional empirical support.

The action-wave mechanics and the reconstructed quantum mechanics clear all
aspects of the EPR problem. RQM shows that the particle states are always
separable and the notion of entanglement of particle states is an ensemble
concept. The correlation of the action-waves and their local interference
reproduces the exact quantum correlations. Einstein locality is strictly
respected. Measurement reveals only the result of local interference of
action-waves. Individually taken, these results are random because the
action-waves corresponding to each particle have an initial random phase.
However, the sum of their phases are fixed by a conservation constraint \ and
this random phase is irrelevant for the correlations. The entangled
wavefunction of standard quantum mechanics does not refer to a single pair of
particles. Naturally it is not a faithful representation of the factual
physical state of the pair of particles. This `incompleteness' is due to the
ensemble averaged nature of the wavefunction and not because of any feature of
quantum dynamics per se. The summary is that with a complete picture of
universal dynamics contained in the Hamiltonian action-wave equation and the
ensemble averaged Schr\"{o}dinger equation, there is nothing to add or change
in quantum mechanics -- as the statistical description of microscopic
dynamics, and only as that, the theory is complete.

I will conclude this appendix with a comment. J. S. Bell derived the maximum
correlation possible in any classical statistical theory that respects
Einstein locality, with hidden stochastic variables and definite prior values
for the observables of a particle \cite{Bell}. Bell found an upper bound for
the spin correlation of two spin-1/2 particles with total spin zero. The
quantum theoretical result for the same correlation in the spin singlet state
exceeds this upper bound. \emph{This implies that quantum theory is not a
classical statistical theory }\cite{Unni-EPL}. (We knew this already;
otherwise Heisenberg would not have written quantum mechanics in terms of
non-commuting relations). There is no information about \emph{Einstein
locality } in the experimental tests; they just measure the correlation and
show that it exceeds the upper bound set for the classical statistical
theories with hidden variables. There is an unfounded impression that the EPR
argued for completing quantum mechanics by replacing it with a classical
statistical theory with hidden variables, of the kind considered by Bell. The
EPR paper does not contain phrases like hidden variables, statistical theory
etc., as can be easily checked. Any such impression is gross distortion of the
content of EPR paper and Einstein's views. Besides, it has been shown that the
local hidden variable theories of the kind Bell considered are incompatible
with the fundamental conservation laws; hence they are unphysical theories
\cite{Unni-EPL,Unni-Pramana}. This result was unfortunately not known when
Bell's theorem was widely discussed. Thus, the experimental tests were about
the viability of theories that were incompatible with the central conservation
laws of physics! In any case, with the new Hamiltonian action-wave mechanics
and RQM, all discussions on completing quantum mechanics can rest.

\end{document}